\RequirePackage{fix-cm}
\RequirePackage{fixltx2e}
\documentclass[12pt,vlined,ruled]{article}
\usepackage[scaled=0.92]{helvet}
\usepackage[T1]{fontenc}
\usepackage[latin9]{inputenc}
\usepackage{geometry}
\usepackage{color}
\usepackage{array}
\usepackage{verbatim}
\usepackage{float}
\usepackage{booktabs}
\usepackage{multirow}
\usepackage{algorithm2e}
\usepackage{amsmath}
\usepackage{amsthm}
\usepackage{amssymb}
\usepackage{graphicx}
\usepackage{setspace}
\usepackage[authoryear]{natbib}
\PassOptionsToPackage{normalem}{ulem}
\usepackage{ulem}
\makeatletter

\providecommand{\tabularnewline}{\\}

  \theoremstyle{definition}
  \newtheorem{defn}{\protect\definitionname}
\theoremstyle{plain}
\newtheorem{thm}{\protect\theoremname}
 \theoremstyle{definition}
  \newtheorem{example}{\protect\examplename}

\RestyleAlgo{boxruled}

\usepackage{titlesec}
\titleformat{\section}[block]{\bfseries\filcenter}{\thesection.}{1em}{}
\titleformat{\subsection}[hang]{\bfseries}{\thesubsection.}{1em}{}

\makeatother

  \providecommand{\definitionname}{Definition}
  \providecommand{\examplename}{Example}
\providecommand{\theoremname}{Theorem}

\begin{document}
\global\long\def\az{\mathcal{A}^{0}}
\global\long\def\as{\mathcal{A}^{*}}
\global\long\def\ak{\mathcal{A}^{k}}
\global\long\def\bs{\mathbb{S}}
\global\long\def\ddd{,\ldots,}
\global\long\def\ba{\mathcal{\boldsymbol{A}}}
\global\long\def\bo{\boldsymbol{\omega}}
\global\long\def\aa{\mathcal{A}}

\title{\textbf{Performance Assessment of High-dimensional Variable Identification}}

\author{Yanjia Yu\thanks{Corresponding author. School of Statistics, University of Minnesota (yuxxx748@umn.edu)},
Yi Yang\thanks{Department of Mathematics and Statistics, McGill University (yi.yang6@mcgill.ca)}
and Yuhong Yang\thanks{School of Statistics, University of Minnesota (yangx374@umn.edu)}}

\date{\today}

\maketitle

\begin{abstract}
Since model selection is ubiquitous in data analysis, reproducibility
of statistical results demands a serious evaluation of reliability
of the employed model selection method, no matter what label it may
have in terms of good properties. Instability measures have been proposed
for evaluating model selection uncertainty. However, low instability
does not necessarily indicate that the selected model is trustworthy,
since low instability can also arise when a certain method tends to
select an overly parsimonious model. $F$- and $G$-measures have
become increasingly popular for assessing variable selection performance
in theoretical studies and simulation results. However, they are not
computable in practice. In this work, we propose an estimation method
for $F$- and $G$-measures and prove their desirable properties of
uniform consistency. This gives the data analyst a valuable tool to
compare different variable selection methods based on the data at
hand. Extensive simulations are conducted to show the very good finite
sample performance of our approach. We further demonstrate the application
of our methods using several micro-array gene expression data sets,
with intriguing findings.
\end{abstract}

\section{INTRODUCTION}

Variable selection in regression and classification is of interest
in many fields, such as bioinformatics, genomics, finance and economics,
etc. For example, in bioinformatics, micro-array gene expression data
are collected to identify cancer-related biomarkers in order to differentiate
affected patients from healthy individuals based on their micro-array
gene expression profile. Typically, the dimension of variables, $p,$
in micro-array gene expression data is of $10^{3-5}$ magnitude, while
the number of subjects, $n,$ is of $10^{1-3}$ magnitude (e.g., \citealp{ma2008penalized}).
For such kind of problems with $p\gg n$, the penalized likelihood
estimation yields a group of methods for selecting a subset of variables
(e.g., \citealp{fan2010selective}). However, it is well recognized
in literature that model selection methods, including the high-dimensional
penalization methods, often encounter variable selection instability
issues \citep{CC1995,DD1995,BL96a,BL96b,BBA97,YZ2005,lim2015estimation}.
For example, removing a few observations or adding small perturbations
to the data may result in dramatically different variable selection
results \citep{meinshausen2006high,chen2007model,nanying,lim2015estimation}.
Clearly, unstable variable selection may have severe practical consequences
in applications. At a larger scale, reproducibility is a major problem
in the science community \citep{mcnutt2014raising,stodden2015reproducing}. 

Previously, variable selection uncertainty is mainly evaluated by
instability measures in the existing literature, which test how sensitive
a variable selection method is to induced small changes of the
data, either by subsampling \citep{chen2007model}, resampling \citep{DE83,BL96b,BBA97}
or adding perturbations \citep{BL96b}. However, low instability measures
do not necessarily indicate that the variable selection results are
reliable, since low instability can also arise when a method tends
to select an overly parsimonious model (e.g. the intercept only model
in the extreme case). 

Therefore, there is a great need for measures that can directly evaluate
the variable selection uncertainty beyond stability. For the purpose
of variable selection, naturally one cares about both types of errors:
including unnecessary variables and excluding important ones. To\textbf{
}summarize the overall performance, $F$- and $G$-measures, often
seen in the field of information retrieval \citep{Chinchor1992MUC4,billsus1998learning},
are becoming very popular for assessing the variable selection performance
(e.g., \citealt{lim2011modeling,lim2015estimation}). Specifically,
$F$-measure is the harmonic mean of\emph{ precision} and \emph{recall},
where precision (or positive predictive value) is defined as the fraction
of selected variables that are true variables, and recall (also known
as sensitivity) is defined as the fraction of the true variables that
are selected. $G$-measure is the geometric mean of precision and
recall \citep{steinbach2000comparison}.

By combining precision and recall into one measure, one can evaluate
the overall accuracy of a given variable selection method. Clearly,
a higher $F$ (or $G$) value indicates better selection performance
in an overall sense. However, previous work in the literature only
calculates the $F$-measure of a given selection method for simulated
data where the true model is known, which cannot be done for real
data.

In this paper, we propose\textbf{ }a method for \textbf{p}erformance
\textbf{a}ssessment of (high-dimensional) \textbf{v}ariable \textbf{i}ndentification
(PAVI) by a combined $F$ or $G$ estimate based on some candidate
models with a proper weighting. Our proposal supports both regression
and classification cases. We provide theoretical justification that
under some sensible conditions, our estimates are uniformly consistent
in estimating the true $F$- and $G$-measures for any set of models
to be checked. The choices of candidate models are very flexible,
which can be obtained by using penalized methods such as Lasso \citep{tibshirani1996regression},
SCAD \citep{fan2001variable}, adaptive Lasso \citep{ZH06}, MCP \citep{zhang2010}
or other variable selection methods. Two weighting methods are considered
in this work: the adaptive regression by mixing \citep{YY01} and weighting
via some information criteria (e.g., \citealp{nanying}). In the simulation
section, we show a very reliable estimation performance of our method
for both classification and regression data. We demonstrate our methods
further by analyzing several micro-array gene expression data. The
real data analysis suggests that PAVI is a very useful tool for evaluating
the variable selection performance of high-dimensional linear based
models. They provide useful information on the reliability and reproducibility
of a given model when the true model is unknown. For example, one
may justifiably doubt the reproducibility of a model that has very
small estimated $F$ and $G$ values.

The remainder of this paper is organized as follows. In Section 2,
we recall the concepts of $F$- and $G$-measures and introduce our
estimation methods. Section 3 provides the theoretical justification
for the $F$- and $G$-measure estimators by PAVI. Section 4 gives
some implementation details for both regression and classification
cases, including how to obtain the candidate models and assign weights.
Simulation results are presented in Section 5. We demonstrate our
methods by analyzing three well-studied gene expression datasets in Section
6. Conclusions are given in Section 7. The technical proofs are relegated
to the Appendix.

\section{METHODOLOGY}

In this paper, we adopt the generalized linear model setting. Denote
$\mathbf{X}=(\mathbf{x}_{1}\ddd\mathbf{x}_{n})^{\intercal}$ the $n\times p$
design matrix with $\mathbf{x}_{i}=(x_{i1}\ddd x_{ip})^{\intercal}$,
$i=1\ddd n$. Let $\mathbf{y}=(y_{1}\ddd y_{n})^{\intercal}$ be the
$n$-dimensional response vector. For regression with a continuous
response, we consider the linear regression model, 
\[
\mathbf{y}=\mathbf{X}\boldsymbol{\beta}^{*}+\boldsymbol{\varepsilon},
\]
where $\boldsymbol{\varepsilon}$ is the vector of $n$ independent
errors and $\boldsymbol{\beta}^{*}=(\beta_{1}^{*}\ddd\beta_{p}^{*})^{\intercal}$
is a $p$-dimensional coefficient vector of the true underlying model
that generates the data. For classification, we use the binary logistic
regression model. Let $Y\in\{0,1\}$ be a binary response variable
and $X\in\mathbb{R}^{p}$ be a $p$-dimensional predictor vector.
We assume that $Y$ follows a Bernoulli distribution given $X=\mathbf{x}$,
with conditional probability
\begin{align}
\mathrm{Pr}(Y=1|X=\mathbf{x}) & =1-\mathrm{Pr}(Y=0|X=\mathbf{x})=\frac{e^{\mathbf{x}^{\intercal}\boldsymbol{\beta^{*}}}}{1+e^{\mathbf{x}^{\intercal}\boldsymbol{\beta^{*}}}}.\label{eq:logit1}
\end{align}
Let $\mathcal{A}^{*}\equiv\mathrm{supp(\boldsymbol{\beta}^{*})}=\{j:\beta_{j}^{*}\neq0\}$
be the index set of the variables in the true model with size $|\as|$,
where $|\cdot|$ denotes the cardinality of a set. For regression
and classification, we assume that the true model is sparse. In other
words, most true coefficients $\beta_{j}^{*}$ in $\boldsymbol{\beta}^{*}$
are exactly zero, except those in $\mathcal{A}^{*}$, i.e. $|\mathcal{A}^{*}|$
is small. 

Let $\az=\{j:\beta_{j}^{0}\neq0\}$ be an index set of all nonzero
coefficients from any given variable selection result. We can use
$F$- and $G$- measures to evaluate the performance of $\az$. $F$-
and $G$-measures take values between 0 and 1, and a higher value
indicates better performance of the variable selection method. The
definitions of $F$- and $G$- measures rely on two quantities, \emph{precision}
and \emph{recall.} The precision $pr$ for $\az$ is the fraction
of true variables in the given model $\az$, i.e. $pr(\az)\equiv pr(\az;\as)=|\az\cap\as|/|\az|$,
and the recall $re$ for $\az$ is the fraction of variables in the
true model $\as$ that are selected, i.e. $re(\az)\equiv re(\az;\as)=|\az\cap\as|/|\as|$.
With the definition of precision and recall, $F$-measure for a given
model $\az$ is defined as the harmonic mean of precision and recall,
while $G$-measure is defined as the geometric mean of the two. Specifically,
\[
F(\az)\equiv F(\az;\as)=\frac{2\times pr(\az)\times re(\az)}{pr(\az)+re(\az)}=\frac{2|\az\cap\as|}{|\az|+|\as|},
\]
and 
\[
G(\az)\equiv G(\az;\as)=\sqrt{pr(\az)\times re(\az)}=\frac{|\az\cap\as|}{\sqrt{|\az|\cdot|\as|}}.
\]

As we know, increasing the regularization level in penalized regression
results in fewer non-zero coefficients, thus fewer active variables
are selected. Therefore, false positives are less likely to happen,
while false negatives become more likely. By taking the harmonic mean
(or geometric mean) of precision and recall, $F$-measure (or $G$-measure)
integrates both false positive and false negative aspects into a single
characterization. For a given $\az$, high $F$- or $G$-measure indicates
that both false positive and false negative rates are low. For example, if
$\as=(1,1,1,0,0,0,0)$ and $\mathcal{A}_{1}^{0}=(1,1,1,0,0,0,1)$,
then $pr(\az_{1})=3/4$, $re(\az_{1})=1$, $F(\az_{1})=6/7$ and $G(\az_{1})=\sqrt{3}/2$. For the same true model $\as,$ if we consider a worse
case where $\mathcal{A}_{2}^{0}=(1,1,0,0,0,0,1)$, then $pr(\az_{2})=2/3$,
$re(\az_{2})=2/3$, $F(\az_{2})=2/3$ and $G(\az_{2})=2/3$. The $F$-
and $G$-measures are smaller than those in the first case due to
the existence of both under-selection and over-selection. In general,
$F$- and $G$-measures are conservative in the sense that both are
more sensitive to under-selection than to over-selection. Specifically,
suppose $|\as|=m$, if $\az_{3}$ over-selects one variable, then
$|\az_{3}|=m+1,$ $F(\az_{3})=2m/(2m+1),$ and $G(\az_{3})=\sqrt{m/(m+1)}$;
if $\az_{4}$ under-selects one variable, then $|\az_{4}|=m-1,$ $F(\az_{4})=(2m-2)/(2m-1),$
and $G(\az_{4})=\sqrt{(m-1)/m}.$ One can easily see that $F(\az_{3})>F(\az_{4})$
and $G(\az_{3})>G(\az_{4}).$

In real applications, the true model $\as$ is usually unknown, and
thus we cannot directly know $F(\az)$ and $G(\az)$ for any given
model $\az$. However, by borrowing information from a group of given
models, we can estimate $F(\az)$ and $G(\az)$ from the data. Suppose
that we have a set of candidate models $\mathbb{\mathbb{S}}=\{\mathcal{A}^{1}\ddd\aa^{K}\}$,
which can be obtained from a preliminary analysis. When the model
size $p$ is small, we can use a full collection of all-subset models
$\mathbb{S=\mathbb{C}}$, where 
\[\mathbb{C}=\{\varnothing,\{1\}\ddd\{p\},\{1,2\},\{1,3\}\ddd\{1\ddd p\}\}\]
with $1\ddd p$ represent the index for $p$ variables. If $p$ is
too large, we can choose $\mathbb{\mathbb{S}}$ as a group of models
obtained from penalized variable selection methods such as Lasso,
adaptive Lasso, SCAD and MCP etc. Define $\mathbf{w}=\{w_{1}\ddd w_{K}\}$
as the corresponding data-driven weights for $\bs=\{\mathcal{A}^{1}\ddd\aa^{K}\}$.
In Section \ref{subsec:Dictionary-models} we will further discuss
how we acquire $\mathbb{S}$ and $\mathbf{w}$. But for now let us
assume they are already properly acquired. For each $\ak$, we define the estimated precision
and recall for $\az$ as $pr(\az;\ak)=|\az\cap\ak|/|\az|$ and $re(\az;\ak)=|\az\cap\ak|/|\ak|$, then we propose the following $\widehat{F}$ by PAVI to estimate $F(\az)$
\begin{equation}
\widehat{F}(\mathcal{A}^{0})=\sum w_{k}F(\az;\ak)=2\sum w_{k}\frac{|\az\cap\ak|}{|\az|+|\ak|}.\label{eq:eq1}
\end{equation}
Similarly, we propose $\widehat{G}$ by PAVI to estimate $G(\az)$
\begin{equation}
\widehat{G}(\mathcal{A}^{0})=\sum w_{k}G(\az;\ak)=2\sum w_{k}\frac{|\az\cap\ak|}{\sqrt{|\az|\cdot|\ak|}}.\label{eq:eq2}
\end{equation}
And we define the standard deviation of $\widehat{F}(\mathcal{A}^{0})$
as
\begin{equation}
\mathrm{sd}\big(\widehat{F}(\mathcal{A}^{0})\big)=\sqrt{\sum w_{k}\big(F(\az;\ak)-\widehat{F}(\mathcal{A}^{0})\big)^{2}}.\label{eq:eq1-1}
\end{equation}
Similarly, the standard deviation of $\widehat{G}(\mathcal{A}^{0})$
is
\begin{equation}
\mathrm{sd}\big(\widehat{G}(\mathcal{A}^{0})\big)=\sqrt{\sum w_{k}\big(G(\az;\ak)-\widehat{G}(\mathcal{A}^{0})\big)^{2}}.\label{eq:eq1-1-1}
\end{equation}

In \eqref{eq:eq1} and \eqref{eq:eq2}, $\widehat{F}(\mathcal{A}^{0})$
and $\widehat{G}(\mathcal{A}^{0})$ are estimated using the candidate
models $\ak\in\mathbb{\bs}$ and weights $w_{k}\in\mathbf{w}$ for
$k=1\ddd K$. Intuitively, if higher weights $w_{k}$'s are assigned
to those $\ak$'s that are close to the true model $\as$, then $\widehat{F}(\mathcal{A}^{0})$
and $\widehat{G}(\mathcal{A}^{0})$ should be able to well approximate
the true values of $F(\az)$ and $G(\az)$ respectively. In Section
\ref{subsec:Estimating-the-weights} we will discuss the methods for
computing weights $\mathbf{w}$ from the data.

\begin{comment}
Notably, there is a connection between the estimated $\widehat{F}(\mathcal{A}^{0})$
and a weighted version of the VSD
\begin{eqnarray*}
1-\widehat{F}(\mathcal{A}^{0}) & = & \sum w_{k}-\sum w_{k}\widehat{F}_{k}=\sum w_{k}\frac{|\az\nabla\ak|}{|\az|+|\ak|}.
\end{eqnarray*}
\end{comment}

\section{THEORY}

In this section, we show that the proposed estimators $\widehat{F}$
and $\widehat{G}$ are uniformly consistent estimators for the true
$F$ and $G$ over the set of all models to be checked. The theory
to be established relies on the property of the data-dependent model
weights $\mathbf{w}=\{w_{1}\ddd w_{K}\}$ referred to as \emph{weak
consistency} \citep{nanying}:
\begin{defn}
[Weak consistency] The weighting vector $\mathbf{w}=(w_{1}\ddd w_{K})^{\intercal}$
is weakly consistent if 

\[
\frac{\sum_{k=1}^{K}w_{k}\cdot|\ak\nabla\as|}{|\as|}\stackrel{p}{\rightarrow}0\quad\mbox{as}\ n\rightarrow\infty,
\]
where $\nabla$ denotes the symmetric difference between two sets.
\end{defn}

The definition basically says that $\mathbf{w}$ is concentrated enough
around the true model $\as$ so that the weighted deviation $|\ak\nabla\as|$
eventually diminishes relative to the size of the true model. When
the true model is allowed to increase in dimension as $n$ increases,
including the denominator $|\as|$ in the definition makes the condition
more likely to be satisfied. When the true model $\as$ is fixed,
weak consistency implies consistency, i.e. $\sum_{k=1}^{K}w_{k}\cdot|\ak\nabla\as|\stackrel{p}{\rightarrow}0,$
as $n\rightarrow\infty.$

The following theorem shows that under the weak consistency condition,
the estimators $\widehat{F}$ and $\widehat{G}$ are uniformly consistent
(the proof is in the Appendix).

\begin{thm}
[Uniform consistency of $\widehat{F}$ and $\widehat{G}$]\label{thm:f-measure}
Suppose the model weighting $\mathbf{w}$ is weakly consistent. Then
$\widehat{F}$ and $\widehat{G}$ based on PAVI are uniformly consistent
in the sense that
\[
\sup_{\az\in\mathbb{\mathbb{C}}}|\widehat{F}(\mathcal{A}^{0})-F(\az)|\overset{p}{\longrightarrow}0\qquad\text{as\ }n\rightarrow\infty;
\]
\[
\sup_{\az\in\mathbb{C}}|\widehat{G}(\mathcal{A}^{0})-G(\az)|\overset{p}{\longrightarrow}0\qquad\mbox{as}\ n\rightarrow\infty.
\]
\end{thm}
From this theorem we see that if the model weighting mostly focuses
on models that are sensibly around the true model, then our estimated
$\widehat{F}$ and $\widehat{G}$ will be close to the true values.
Clearly, we also have $E|\widehat{F}(\az)-F(\az)|\rightarrow0$ and
$E|\widehat{G}(\az)-G(\az)|\rightarrow0$ uniformly.
\begin{thm}
[Uniform convergence of $\mathrm{sd}\big(\widehat{F}\big)$ and $\mathrm{sd}\big(\widehat{G}\big)$]\label{thm:sd-f-measure}
Suppose the model weighting $\mathbf{w}$ is weakly consistent. Then
$\mathrm{sd}\big(\widehat{F}\big)$ and $\mathrm{sd}\big(\widehat{G}\big)$
based on PAVI converge to 0 in probability uniformly in the sense
that
\[
\sup_{\az\in\mathbb{\mathbb{C}}}|\mathrm{sd}\big(\widehat{F}(\mathcal{A}^{0})\big)|\overset{p}{\longrightarrow}0\qquad\text{as\ }n\rightarrow\infty;
\]
\[
\sup_{\az\in\mathbb{C}}|\mathrm{sd}\big(\widehat{G}(\mathcal{A}^{0})\big)|\overset{p}{\longrightarrow}0\qquad\mbox{as}\ n\rightarrow\infty.
\]
\end{thm}
From this theorem we see that if the model weighting is sensible,
then $\mathrm{sd}\big(\widehat{F}\big)$ and $\mathrm{sd}\big(\widehat{G}\big)$
will be close to 0. The results also support reliability of our PAVI
method.

\section{IMPLEMENTATION}

\subsection{Candidate models\label{subsec:Dictionary-models}}

Now we discuss how to choose the candidate models for computing $\widehat{F}$
and $\widehat{G}$. To get the candidate models, we can use a complete
collection of all-subset models, i.e. choose $\mathbb{S}=\mathbb{C}$.
However, in the high-dimensional case where $p\gg n$, it is almost
impossible to use all-subsets due to high computational cost. Instead
we obtain the candidate models by combining the models on the solution
paths of the high-dimensional penalized generalized linear models.
We show in the following how it is done for the logistic regression
models and similar procedures apply to linear regression models. Given
$n$ independent observations $\{(\mathbf{x}_{i},y_{i})\}_{i=1}^{n}$
for the pair $(X,Y)$, let $\pi_{i}=\mathrm{Pr}(Y_{i}=1|X_{i}=\mathbf{x}_{i})$
be the probability in \eqref{eq:logit1} for observation $i$, then
we can fit the logistic regression model by maximizing the penalized
log-likelihood
\begin{equation}
\max_{\boldsymbol{\beta}\in\mathbb{R}^{p}}\ell(\boldsymbol{\beta})-\sum_{j=1}^{p}p_{\lambda}(\beta_{j}),\label{eq:obj}
\end{equation}
where
\[
\ell(\boldsymbol{\beta})=\frac{1}{n}\sum_{i=1}^{n}\left\{ y_{i}\log\pi_{i}+(1-y_{i})\log(1-\pi_{i})\right\} .
\]
Here the nonnegative penalty function $p_{\lambda}(\cdot)$ with regularization
parameter $\lambda\in(0,\infty)$ can be Lasso \citep{tibshirani1996regression}
with penalty $p_{\lambda}(u)=\lambda|u|$, or non-convex penalties
such as SCAD \citep{fan2001variable} penalty, whose derivative is
given by
\begin{eqnarray*}
p_{\lambda}'(u) & = & \lambda\left\{ I(u\leq\lambda)+\frac{(a\lambda-u)_{+}}{\lambda(a-1)}I(u>\lambda)\right\} \quad(a>2),
\end{eqnarray*}
and the MCP penalty \citep{zhang2010} with the derivative
\[
p'_{\lambda}(u)=\frac{(a\lambda-u)_{+}}{a},\qquad(a>1).
\]
We can compute the models $\bs=\{\aa{}^{\lambda_{1}}\ddd\aa^{\lambda_{L}}\}$
for Lasso, SCAD and MCP respectively on the solution paths $\{\widehat{\boldsymbol{\beta}}^{\lambda_{1}}\ddd\widehat{\boldsymbol{\beta}}^{\lambda_{L}}\}$
for decreasing sequences of tuning parameters $\{\lambda_{1}\ddd\lambda_{L}\}$.
These models are then combined together as a set of candidate models
$\mathbb{\bs}=\{\bs_{\mathrm{Lasso}},\mathbb{\bs}_{\mathrm{SCAD}},\mathbb{\bs}_{\mathrm{MCP}}\}$.
One can efficiently compute the whole solution paths of Lasso using
\textbf{glmnet }algorithm \citep{friedman2010regularization}, and
using \textbf{ncvreg }algorithm \citep{breheny2011coordinate} for
SCAD and MCP.

\subsection{Weighting methods \label{subsec:Estimating-the-weights}}

In this section we discuss several different methods in the literature
for determining the weights $\mathbf{w}=\{w_{1}\ddd w_{K}\}$. For
example, \citet{BBA97} and \citet{LB06} proposed information criterion
based methods for weighting, such as those using AIC \citep{AIC73} and
BIC \citep{BIC78}; \citet{HMR1999} proposed the Bayesian model averaging
(BMA) method for weighting; \citet{YY01} studied a weighting strategy
called the adaptive regression by mixing (ARM), which can be computed
by data splitting and cross-assessment. It was proven in \citet{YY01}
that the weighting by ARM delivers the best rate of convergence for
regression estimation. In \citet{yang2000adaptive}, the ARM weighting
method was also extended to the classification setting. When the number
of models in the candidate-model set is fixed, BMA weighting is consistent
(thus weakly consistent). From \citet{YY2007}, when one properly
chooses the data splitting ratio, the ARM weighting can be consistent. More recently, \citet{fiducial2015} proposed Fisher's fiducial based methods for deriving probability density functions as weights on the set of candidate models. They showed that, under certain conditions, their method is consistent when $p$ is diverging and the size of true model is fixed or diverging. In this paper, we only consider the ARM weighting and weighting based on an information criterion. 

\subsubsection*{Weighting using ARM for logistic regression model}

To get the ARM weights, we randomly split the data $\mathbf{D}=\{(\mathbf{x}_{i},y_{i})\}_{i=1}^{n}$
equally into a training set $\mathbf{D}_{1}$ and a test set $\mathbf{D}_{2}$.
Then the logistic regression model is trained on $\mathbf{D}_{1}$
and its prediction performance is evaluated on $\mathbf{D}_{2}$,
based on which the weights $\mathbf{w}=\{w_{1}\ddd w_{K}\}$ can be
computed. Let $\boldsymbol{\beta}_{s}^{(k)}$ be the sub-vector of
$\boldsymbol{\beta}^{(k)}$ representing the nonzero coefficients
of model $\ak$, and let $\mathbf{x}_{s}^{(k)}\in\mathbb{R}^{|\ak|}$
be the corresponding subset of selected predictors. When $p$ is large,
the ARM weighting performs very poorly for measuring the model deviation.
One way to fix this problem is to add a non-uniform prior $e^{-\psi C_{k}}$
in the weighting computation, where $C_{k}=s_{k}\log{\frac{ep}{s_{k}}}+2\log(s_{k}+2)$
and $s_{k}$ is the number of non-constant predictors for model $k$.
The ARM weighting method is summarized in Algorithm \ref{alg:ARM-weighting-method}.

\begin{algorithm}
\begin{enumerate}
\item {\footnotesize{}Randomly split $\mathbf{D}$ into a training set $\mathbf{D}_{1}$
and a test set $\mathbf{D}_{2}$ of equal size.}{\footnotesize \par}
\item {\footnotesize{}For each $\ak\in\mathbb{\bs}$, fit a standard logistic
regression of $y$ on $\mathbf{x}_{s}^{(k)}$ using the }\\
{\footnotesize{}samples in $\mathbf{D}_{1}$ and get the estimated
conditional probability function $\hat{p}^{(k)}(\mathbf{x}_{s}^{(k)})$,
\begin{align*}
\hat{p}^{(k)}(\mathbf{x}_{s}^{(k)}) & \equiv\mbox{Pr}(Y=1|X_{s}^{(k)}=\mathbf{x}_{s}^{(k)})\\
 & =\exp(\mathbf{x}_{s}^{(k)\intercal}\widehat{\boldsymbol{\beta}}_{s}^{(k)})/(1+\exp(\mathbf{x}_{s}^{(k)\intercal}\widehat{\boldsymbol{\beta}}_{s}^{(k)})),\qquad k=1\ddd K.
\end{align*}
}{\footnotesize \par}
\item {\footnotesize{}For each $\ak$, evaluate $\hat{p}^{(k)}(\mathbf{x}_{s}^{(k)})$
on the test set $\mathbf{D}_{2}$.}{\footnotesize \par}
\item {\footnotesize{}Compute the weight $w_{k}$ for each model $\ak$
in the candidate models:}

{\footnotesize \par}{\footnotesize{}
\[
w_{k}=\frac{e^{-\psi C_{k}}\prod_{(\mathbf{x}_{s,i}^{(k)},y_{i})\in\mathbf{D}_{2}}\hat{p}^{(k)}(\mathbf{x}_{s,i}^{(k)})^{y_{i}}\left(1-\hat{p}^{(k)}(\mathbf{x}_{s,i}^{(k)})\right)^{1-y_{i}}}{\sum_{l=1}^{K}e^{-\psi C_{l}}\prod_{(\mathbf{x}_{s,i}^{(l)},y_{i})\in\mathbf{D}_{2}}\hat{p}^{(l)}(\mathbf{x}_{s,i}^{(l)})^{y_{i}}\left(1-\hat{p}^{(l)}(\mathbf{x}_{s,i}^{(l)})\right)^{1-y_{i}}},\ k=1\ddd K.
\]
}{\footnotesize \par}
\item {\footnotesize{}Repeat the steps above (with random data splitting)
$L$ times to get $w_{k}^{(l)}$ }\\
{\footnotesize{}for $l=1\ddd L$, and get $w_{k}=\frac{1}{L}\sum_{l=1}^{L}w_{k}^{(l)}$.}{\footnotesize \par}
\end{enumerate}
\caption{The procedure for the ARM weighting in the classification case.\label{alg:ARM-weighting-method}}
\end{algorithm}

\subsubsection*{Weighting using ARM for linear regression model}

The ARM weighting for the linear regression model 
\[
y_{i}=\mathbf{x}_{i}^{\top}\boldsymbol{\beta}^{*}+\epsilon_{i},\quad\epsilon_{i}\sim N(0,\sigma^{2})
\]
is described in Algorithm \ref{alg:ARM-weighting-method-reg}. 

\begin{algorithm}
\begin{enumerate}
\item {\footnotesize{}Randomly split $\mathbf{D}$ into a training set $\mathbf{D}_{1}$
and a test set $\mathbf{D}_{2}$ of equal size.}{\footnotesize \par}
\item {\footnotesize{}For each $\ak\in\mathbb{\bs}$, fit a standard linear
regression of $y$ on $\mathbf{x}_{s}^{(k)}$ using the training }\\
{\footnotesize{}set $\mathbf{D}_{1}$ and get the estimated regression
coefficient $\widehat{\boldsymbol{\beta}}_{s}^{(k)}$ and the estimated
}\\
{\footnotesize{}standard deviation $\widehat{\boldsymbol{\sigma}}_{s}^{(k)}$.}{\footnotesize \par}
\item {\footnotesize{}For each $\ak$, compute the prediction $\mathbf{x}_{s}^{(k)\intercal}\widehat{\boldsymbol{\beta}}_{s}^{(k)}$
on the test set $\mathbf{D}_{2}$.}{\footnotesize \par}
\item {\footnotesize{}Compute the weight $w_{k}$ for each candidate model
$\ak$:}

{\footnotesize \par}{\footnotesize{}
\[
w_{k}=\frac{e^{-\psi C_{k}}(\widehat{\boldsymbol{\sigma}}_{s}^{(k)})^{-n/2}\prod_{(\mathbf{x}_{si}^{(k)},y_{i})\in\mathbf{D}_{2}}\exp(-(\widehat{\boldsymbol{\sigma}}_{s}^{(k)})^{-2}(y_{i}-\mathbf{x}_{s}^{(k)\intercal}\widehat{\boldsymbol{\beta}}_{s}^{(k)})^{2}/2)}{\sum_{l=1}^{K}e^{-\psi C_{l}}(\widehat{\boldsymbol{\sigma}}_{s}^{(l)})^{-n/2}\prod_{(\mathbf{x}_{si}^{(l)},y_{i})\in\mathbf{D}_{2}}\exp(-(\widehat{\boldsymbol{\sigma}}_{s}^{(l)})^{-2}(y_{i}-\mathbf{x}_{s}^{(l)\intercal}\widehat{\boldsymbol{\beta}}_{s}^{(k)})^{2}/2)},
\]
for $k=1\ddd K$, where $C_{k}=s_{k}\log{\frac{e\cdot p}{s_{k}}}+2\log(s_{k}+2),~k=1,\dots,K$. }{\footnotesize \par}
\item {\footnotesize{}Repeat the steps above (with random data splitting)
$L$ times to get $w_{k}^{(l)}$ }\\
{\footnotesize{}for $l=1\ddd L$, and get $w_{k}=\frac{1}{L}\sum_{l=1}^{L}w_{k}^{(l)}$.}{\footnotesize \par}
\end{enumerate}
\caption{The procedure for the ARM weighting in the regression case.\label{alg:ARM-weighting-method-reg}}
\end{algorithm}

\subsubsection*{Weighting using modified BIC for logistic regression model and linear
regression model}

Information criteria such as BIC can be used as alternative ways for
computing weights. Let $\ell_{k}$ be the maximized likelihood. Recall
that BIC is given by $I_{k}^{\mathrm{BIC}}=-2\log\ell_{k}+s_{k}\log n$.
To accommodate the huge number of models, an extra term was added
by \citet{yang1998asymptotic} to reflect the additional price we
need to pay for searching through all the models. Including the extra
term in the information criteria, we calculate the weights by using
a modified BIC (BIC-p) information criterion:
\begin{equation}
w_{k}=\exp(-I_{k}/2-\psi C_{k})/\sum_{l=1}^{K}\exp(-I_{l}/2-\psi C_{l}),\ k=1\ddd K,\label{eq:AICBIC}
\end{equation}
where $C_{k}=s_{k}\log{\frac{ep}{s_{k}}}+2\log(s_{k}+2)$. 

\section{SIMULATION\label{sec:Simulation}}

In this section, in order to study the performance of estimated $F$-
and $G$-measures, we conduct simulations for several well-known variable
selection methods (for both regression and classification models)
under various settings. We consider numerical experiments for both
$n<p$ and $n>p$ cases, with specified structural feature correlation
(independent/correlated). We also consider some special settings of
the true coefficients such as decaying coefficients. 

\subsection{Setting I: classification models}

For the classification case, we randomly generate $n$ i.i.d observations
$\{y_{i},\mathbf{x}_{i}\}_{i=1}^{n}$. Each binary response $y_{i}\in\{0,1\}$
is generated according to the Bernoulli distribution with the conditional
probability $\mathrm{Pr}(Y=1|X=\mathbf{x}_{i})=1-\mathrm{Pr}(Y=0|X=\mathbf{x}_{i})=\frac{\exp(\mathbf{x}_{i}^{\intercal}\boldsymbol{\beta})}{1+\exp(\mathbf{x}_{i}^{\intercal}\boldsymbol{\beta})}$.
The predictors $\mathbf{x}_{i}$ and the coefficient vector $\boldsymbol{\beta}$
are generated according to the following settings:
\begin{example}
\label{eg:fg-1} $n=200$, $p=8$, $\boldsymbol{\beta}=(3,1.5,2,0,0,0,0,0)^{\intercal}.$
Predictors $\mathbf{x}_{i}$ for $i=1\ddd n$ are generated as $n$
i.i.d. observations from $N_{p}(0,\boldsymbol{I})$. 
\end{example}

\begin{example}
\label{eg:fg-2} Same as Example 1 except $n=1000$.
\end{example}

\begin{example}
\label{eg:fg-3} $n=200$, $p=2000$, $\boldsymbol{\beta}=(\beta_{1},\ldots,\beta_{p})^{\intercal}$,
where $(\beta_{1},\beta_{2},\beta_{3})=(3,1.5,2)$ and $(\beta_{4},\ldots,\beta_{2000})$
are zeros. Predictors $\mathbf{x}_{i}$ for $i=1\ddd n$ are sampled
as $n$ i.i.d. observations from $N(0,\boldsymbol{I}_{p})$.
\end{example}

\begin{example}
\label{eg:fg-4} $n=200$, $p=30$, the components 1--5
of $\boldsymbol{\beta}$ are 10.5, components 6--10 are
5.5, components 11--15 are 0.5 and the rests are zeros.
So there are 15 nonzero predictors, including five large ones, five
moderate ones and five small ones. Predictors $\mathbf{x}_{i}$ for
$i=1\ddd n$ are generated from $X\sim N_{p}(0,\boldsymbol{\Sigma})$
with $\boldsymbol{\Sigma}=(0.4^{|j-k|})_{p\times p}$, thus the pairwise
correlation between $X_{j}$ and $X_{k}$ is $0.4^{|j-k|}$.
\end{example}

\begin{example}
\label{eg:fg-5} $n=200$, $p=200$, the components 1--5
of $\beta$ are 10.5, the components 6--10 are 5.5, the
components 11--15 are 0.5 and the rests are zeros. Predictors
$\mathbf{x}_{i}$ for $i=1\ddd n$ are generated from $X\sim N_{p}(0,\boldsymbol{\Sigma})$.
The covariance structure $\boldsymbol{\Sigma}$ is set as follows:
the first 15 predictors $(X_{1},\ldots,X_{15})$ and the remaining
185 predictors $(X_{16},\ldots,X_{200})$ are independent. The pairwise
correlation between $X_{j}$ and $X_{k}$ in $(X_{1},\ldots,X_{15})$
is $0.4^{|j-k|}$ with $j,k=1,\ldots,15$. The pairwise correlation
between $X_{j}$ and $X_{k}$ in $(X_{16},\ldots,X_{200})$ is $0.4^{|j-k|}$
with $j,k=16,\ldots,200$.
\end{example}

We fit four penalized methods, Lasso, adaptive Lasso, MCP and SCAD
on the data from Examples 1--5, and denoted by $\mathcal{A}^{\mathrm{Lasso}}$,
$\mathcal{A}^{\mathrm{AdLasso}}$, $\mathcal{A}^{\mathrm{MCP}}$ and
$\mathcal{A}^{\mathrm{SCAD}}$ the resulting models respectively.
The \textbf{glmnet} algorithm \citep{friedman2010regularization}
is used for computing $\mathcal{A}^{\mathrm{Lasso}}$ and $\mathcal{A}^{\mathrm{AdLasso}}$,
and \textbf{ncvreg} \citep{breheny2011coordinate} is used for computing
$\mathcal{A}^{\mathrm{MCP}}$ and $\mathcal{A}^{\mathrm{SCAD}}$.
Five-fold cross-validation is used for penalization parameter tuning
for those procedures. Because we know the true model $\mathcal{A}^{*}=\{j:\beta_{j}\neq0\}$
in the simulation, we can report the true $F(\az)$ and $G(\az)$
measures for each model-under-check $\az\in\{\mathcal{A}^{\mathrm{Lasso}}$,
$\mathcal{A}^{\mathrm{AdLasso}}$, $\mathcal{A}^{\mathrm{MCP}}$,
$\mathcal{A}^{\mathrm{SCAD}}$\}. For comparison, we also compute
estimated $\widehat{F}$ and $\widehat{G}$ using two different weighting
methods, ARM and BIC-p (the modified BIC) with prior adjustment $\psi=1$.
The absolute differences between the true measures and the estimated
measures are used to measure estimation performances, i.e. 
\begin{align*}
d_{F} & =|\widehat{F}(\mathcal{A}^{0})-F(\az)|,\\
d_{G} & =|\widehat{G}(\mathcal{A}^{0})-G(\az)|,
\end{align*}
where the smaller $d_{F}$ and $d_{G}$ values indicate better estimation
performance. The number of observations in the training set for computing
the ARM weight is half of the sample size $\left\lfloor n/2\right\rfloor $,
and the corresponding repetition number is 100. 

All simulation examples are repeated for 100 times and the corresponding
$F(\az)$, $G(\az)$, $\widehat{F}(\mathcal{A}^{0})$, $\widehat{G}(\mathcal{A}^{0})$,
$d_{F}$ and $d_{G}$ values are computed and averaged. The results
are summarized in Tables \ref{tab:example-1}--\ref{tab:example-5}.
The standard errors are also shown in parentheses. As we can see in
those tables, $d_{F}$ and $d_{G}$ are generally small, which indicates
that the estimated $\widehat{F}(\mathcal{A}^{0})$ and $\widehat{G}(\mathcal{A}^{0})$
are good approximations to the true $F(\az)$ and $G(\az)$. The estimated
$\widehat{F}(\mathcal{A}^{0})$ and $\widehat{G}(\mathcal{A}^{0})$
can reflect the true advantage of a given variable selection method.
For example, in Tables \ref{tab:example-1}--\ref{tab:example-5},
we can see that adaptive Lasso, MCP and SCAD have better variable
selection performance than Lasso according to their larger true $F(\az)$
and $G(\az)$. The estimated $\widehat{F}(\mathcal{A}^{0})$ and $\widehat{G}(\mathcal{A}^{0})$
can correctly reflect these performance differences. 

Our estimation method can still perform very well under the high-dimensional
setting, which can be seen from the small $d_{F}$ and $d_{G}$ in
Table \ref{tab:example-3}. However, the results from Tables 4 and
5 show that the decaying coefficients and feature correlation make
the estimation of $\widehat{F}(\mathcal{A}^{0})$ and $\widehat{G}(\mathcal{A}^{0})$
more difficult. In those two cases, BIC-p methods tend to over-estimate
$F(\az)$ and $G(\az)$ for MCP and SCAD models, while ARM tends to
under-estimate $F(\az)$ and $G(\az)$ for Lasso and adaptive Lasso. 

The overestimation problem of the BIC-p method mainly comes from overestimation
of the recall part. The final model selected by SCAD misses several true
variables, thus the true recall is very small. However, if one uses
the heavily weighted candidate models that miss several true variables
in the PAVI calculation, the recall would be overestimated. 

For SCAD and ARM combination, using the heavily weighted models that
miss several true variables in PAVI will give us over-estimation of
the recall and under-estimation of precision, while these two effects
cancel each other to some degree.

The underestimation by ARM methods mainly comes from the underestimation
of the precision part, while the estimated recall is close (slightly
overestimation) to the true recall. Lasso tends to miss true variables
and over-select redundant variables in the example. Thus, the true
precision of Lasso is small. However, if one uses the heavily weighted
candidate models in PAVI for true model, Lasso's over-selection appears
to be more severe. So the precision would be underestimated. 

For Lasso and BIC combination, using the heavily weighted models that
miss several true variables with small coefficients in PAVI computing
will give us over-estimation of the recall and under-estimation of
precision, while these two effects cancel each other to some degree. 

Both issues are mainly caused by the fact that the candidate models with large
weights could not recover all the variables with small true coefficients,
and the problem is further worsened by the existence of high feature
correlation. 

\begin{table}
\caption{Classification case (Example 1): $n=200$, $p=8$, $\boldsymbol{\beta}=(3,1.5,2,0,0,0,0,0)^{\intercal}.$
$\mathbf{x}_{i}$ for $i=1\protect\ddd n$ are i.i.d. from $N_{p}(0,\boldsymbol{I})$.
Values are averaged over 100 independent runs. The standard errors
are shown in parentheses. \label{tab:example-1}}
\begin{centering}
{\footnotesize{}}%
\begin{tabular}{lr@{\extracolsep{0pt}.}lr@{\extracolsep{0pt}.}lr@{\extracolsep{0pt}.}lr@{\extracolsep{0pt}.}l}
\toprule 
 & \multicolumn{2}{c}{{\footnotesize{}$F$}} & \multicolumn{2}{c}{{\footnotesize{}$G$}} & \multicolumn{2}{c}{{\footnotesize{}$d_{F}$}} & \multicolumn{2}{c}{{\footnotesize{}$d_{G}$}}\tabularnewline
\midrule 
\multicolumn{9}{c}{{\footnotesize{}Lasso}}\tabularnewline
{\footnotesize{}True} & {\footnotesize{}0}&{\footnotesize{}670 (0.010)} & {\footnotesize{}0}&{\footnotesize{}712 (0.009)} & \multicolumn{2}{c}{} & \multicolumn{2}{c}{}\tabularnewline
{\footnotesize{}ARM} & {\footnotesize{}0}&{\footnotesize{}711 (0.009)} & {\footnotesize{}0}&{\footnotesize{}747 (0.007)} & {\footnotesize{}0}&{\footnotesize{}046 (0.003)} & {\footnotesize{}0}&{\footnotesize{}039 (0.002)}\tabularnewline
{\footnotesize{}BIC-p} & {\footnotesize{}0}&{\footnotesize{}687 (0.010)} & {\footnotesize{}0}&{\footnotesize{}726 (0.008)} & {\footnotesize{}0}&{\footnotesize{}017 (0.002)} & {\footnotesize{}0}&{\footnotesize{}014 (0.001)}\tabularnewline
\midrule
\multicolumn{9}{c}{{\footnotesize{}AdLasso}}\tabularnewline
{\footnotesize{}True} & {\footnotesize{}0}&{\footnotesize{}944 (0.009)} & {\footnotesize{}0}&{\footnotesize{}949 (0.008)} & \multicolumn{2}{c}{} & \multicolumn{2}{c}{}\tabularnewline
{\footnotesize{}ARM} & {\footnotesize{}0}&{\footnotesize{}899 (0.004)} & {\footnotesize{}0}&{\footnotesize{}908 (0.004)} & {\footnotesize{}0}&{\footnotesize{}066 (0.003)} & {\footnotesize{}0}&{\footnotesize{}060 (0.003)}\tabularnewline
{\footnotesize{}BIC-p} & {\footnotesize{}0}&{\footnotesize{}946 (0.007)} & {\footnotesize{}0}&{\footnotesize{}950 (0.007)} & {\footnotesize{}0}&{\footnotesize{}018 (0.002)} & {\footnotesize{}0}&{\footnotesize{}016 (0.001)}\tabularnewline
\midrule
\multicolumn{9}{c}{{\footnotesize{}MCP}}\tabularnewline
{\footnotesize{}True} & {\footnotesize{}0}&{\footnotesize{}968 (0.009)} & {\footnotesize{}0}&{\footnotesize{}971 (0.008)} & \multicolumn{2}{c}{} & \multicolumn{2}{c}{}\tabularnewline
{\footnotesize{}ARM} & {\footnotesize{}0}&{\footnotesize{}903 (0.005)} & {\footnotesize{}0}&{\footnotesize{}913 (0.004)} & {\footnotesize{}0}&{\footnotesize{}079 (0.003)} & {\footnotesize{}0}&{\footnotesize{}072 (0.002)}\tabularnewline
{\footnotesize{}BIC-p} & {\footnotesize{}0}&{\footnotesize{}961 (0.007)} & {\footnotesize{}0}&{\footnotesize{}965 (0.006)} & {\footnotesize{}0}&{\footnotesize{}019 (0.002)} & {\footnotesize{}0}&{\footnotesize{}017 (0.001)}\tabularnewline
\midrule
\multicolumn{9}{c}{{\footnotesize{}SCAD}}\tabularnewline
{\footnotesize{}True} & {\footnotesize{}0}&{\footnotesize{}902 (0.012)} & {\footnotesize{}0}&{\footnotesize{}911 (0.010)} & \multicolumn{2}{c}{} & \multicolumn{2}{c}{}\tabularnewline
{\footnotesize{}ARM} & {\footnotesize{}0}&{\footnotesize{}881 (0.006)} & {\footnotesize{}0}&{\footnotesize{}892 (0.006)} & {\footnotesize{}0}&{\footnotesize{}054 (0.003)} & {\footnotesize{}0}&{\footnotesize{}050 (0.003)}\tabularnewline
{\footnotesize{}BIC-p} & {\footnotesize{}0}&{\footnotesize{}911 (0.010)} & {\footnotesize{}0}&{\footnotesize{}919 (0.009)} & {\footnotesize{}0}&{\footnotesize{}018 (0.002)} & {\footnotesize{}0}&{\footnotesize{}016 (0.001)}\tabularnewline
\bottomrule
\end{tabular}
\par\end{centering}{\footnotesize \par}

\end{table}

\begin{table}
\caption{Classification case (Example 2): $n=1000$, $p=8$, $\boldsymbol{\beta}=(3,1.5,2,0,0,0,0,0)^{\intercal}.$
$\mathbf{x}_{i}$ for $i=1\protect\ddd n$ are i.i.d. from $N_{p}(0,\boldsymbol{I})$.
Values are averaged over 100 independent runs. The standard errors
are shown in parentheses. \label{tab:example-2}}
\begin{centering}
{\footnotesize{}}%
\begin{tabular}{lr@{\extracolsep{0pt}.}lr@{\extracolsep{0pt}.}lr@{\extracolsep{0pt}.}lr@{\extracolsep{0pt}.}l}
\toprule 
 & \multicolumn{2}{c}{{\footnotesize{}$F$}} & \multicolumn{2}{c}{{\footnotesize{}$G$}} & \multicolumn{2}{c}{{\footnotesize{}$d_{F}$}} & \multicolumn{2}{c}{{\footnotesize{}$d_{G}$}}\tabularnewline
\midrule 
\multicolumn{9}{c}{{\footnotesize{}Lasso}}\tabularnewline
{\footnotesize{}True} & {\footnotesize{}0}&{\footnotesize{}631 (0.008)} & {\footnotesize{}0}&{\footnotesize{}680 (0.006)} & \multicolumn{2}{c}{} & \multicolumn{2}{c}{}\tabularnewline
{\footnotesize{}ARM} & {\footnotesize{}0}&{\footnotesize{}697 (0.007)} & {\footnotesize{}0}&{\footnotesize{}734 (0.006)} & {\footnotesize{}0}&{\footnotesize{}066 (0.002)} & {\footnotesize{}0}&{\footnotesize{}054 (0.002)}\tabularnewline
{\footnotesize{}BIC-p} & {\footnotesize{}0}&{\footnotesize{}639 (0.008)} & {\footnotesize{}0}&{\footnotesize{}686 (0.006)} & {\footnotesize{}0}&{\footnotesize{}008 (0.001)} & {\footnotesize{}0}&{\footnotesize{}006 (0.001)}\tabularnewline
\midrule
\multicolumn{9}{c}{{\footnotesize{}AdLasso}}\tabularnewline
{\footnotesize{}True} & {\footnotesize{}0}&{\footnotesize{}989 (0.004)} & {\footnotesize{}0}&{\footnotesize{}989 (0.004)} & \multicolumn{2}{c}{} & \multicolumn{2}{c}{}\tabularnewline
{\footnotesize{}ARM} & {\footnotesize{}0}&{\footnotesize{}929 (0.002)} & {\footnotesize{}0}&{\footnotesize{}935 (0.002)} & {\footnotesize{}0}&{\footnotesize{}067 (0.002)} & {\footnotesize{}0}&{\footnotesize{}062 (0.002)}\tabularnewline
{\footnotesize{}BIC-p} & {\footnotesize{}0}&{\footnotesize{}987 (0.003)} & {\footnotesize{}0}&{\footnotesize{}988 (0.002)} & {\footnotesize{}0}&{\footnotesize{}009 (0.001)} & {\footnotesize{}0}&{\footnotesize{}008 (0.001)}\tabularnewline
\midrule
\multicolumn{9}{c}{{\footnotesize{}MCP}}\tabularnewline
{\footnotesize{}True} & {\footnotesize{}0}&{\footnotesize{}964 (0.008)} & {\footnotesize{}0}&{\footnotesize{}967 (0.008)} & \multicolumn{2}{c}{} & \multicolumn{2}{c}{}\tabularnewline
{\footnotesize{}ARM} & {\footnotesize{}0}&{\footnotesize{}922 (0.004)} & {\footnotesize{}0}&{\footnotesize{}929 (0.004)} & {\footnotesize{}0}&{\footnotesize{}065 (0.002)} & {\footnotesize{}0}&{\footnotesize{}059 (0.002)}\tabularnewline
{\footnotesize{}BIC-p} & {\footnotesize{}0}&{\footnotesize{}965 (0.008)} & {\footnotesize{}0}&{\footnotesize{}968 (0.007)} & {\footnotesize{}0}&{\footnotesize{}009 (0.001)} & {\footnotesize{}0}&{\footnotesize{}008 (0.001)}\tabularnewline
\midrule
\multicolumn{9}{c}{{\footnotesize{}SCAD}}\tabularnewline
{\footnotesize{}True} & {\footnotesize{}0}&{\footnotesize{}955 (0.010)} & {\footnotesize{}0}&{\footnotesize{}960 (0.009)} & \multicolumn{2}{c}{} & \multicolumn{2}{c}{}\tabularnewline
{\footnotesize{}ARM} & {\footnotesize{}0}&{\footnotesize{}919 (0.005)} & {\footnotesize{}0}&{\footnotesize{}926 (0.004)} & {\footnotesize{}0}&{\footnotesize{}065 (0.002)} & {\footnotesize{}0}&{\footnotesize{}059 (0.002)}\tabularnewline
{\footnotesize{}BIC-p} & {\footnotesize{}0}&{\footnotesize{}956 (0.009)} & {\footnotesize{}0}&{\footnotesize{}961 (0.008)} & {\footnotesize{}0}&{\footnotesize{}009 (0.001)} & {\footnotesize{}0}&{\footnotesize{}008 (0.001)}\tabularnewline
\bottomrule
\end{tabular}
\par\end{centering}{\footnotesize \par}

\end{table}

\begin{table}
\caption{Classification case (Example 3): $n=200$, $p=2000$, $\boldsymbol{\beta}=(\beta_{1},\ldots,\beta_{p})^{\intercal}$,
where $(\beta_{1},\beta_{2},\beta_{3})=(3,1.5,2)$ and $(\beta_{4},\ldots,\beta_{2000})$
are zeros. $\mathbf{x}_{i}$ for $i=1\protect\ddd n$ are i.i.d. from
$N(0,\boldsymbol{I}_{p})$. Values are averaged over 100 independent
runs. The standard errors are shown in parentheses. \label{tab:example-3}}
\begin{centering}
{\footnotesize{}}%
\begin{tabular}{lr@{\extracolsep{0pt}.}lr@{\extracolsep{0pt}.}lr@{\extracolsep{0pt}.}lr@{\extracolsep{0pt}.}l}
\toprule 
 & \multicolumn{2}{c}{{\footnotesize{}$F$}} & \multicolumn{2}{c}{{\footnotesize{}$G$}} & \multicolumn{2}{c}{{\footnotesize{}$d_{F}$}} & \multicolumn{2}{c}{{\footnotesize{}$d_{G}$}}\tabularnewline
\midrule 
\multicolumn{9}{c}{{\footnotesize{}Lasso}}\tabularnewline
{\footnotesize{}True} & {\footnotesize{}0}&{\footnotesize{}154 (0.011)} & {\footnotesize{}0}&{\footnotesize{}278 (0.010)} & \multicolumn{2}{c}{} & \multicolumn{2}{c}{}\tabularnewline
{\footnotesize{}ARM} & {\footnotesize{}0}&{\footnotesize{}129 (0.009)} & {\footnotesize{}0}&{\footnotesize{}251 (0.009)} & {\footnotesize{}0}&{\footnotesize{}025 (0.002)} & {\footnotesize{}0}&{\footnotesize{}028 (0.002)}\tabularnewline
{\footnotesize{}BIC-p} & {\footnotesize{}0}&{\footnotesize{}159 (0.011)} & {\footnotesize{}0}&{\footnotesize{}283 (0.010)} & {\footnotesize{}0}&{\footnotesize{}010 (0.002)} & {\footnotesize{}0}&{\footnotesize{}010 (0.002)}\tabularnewline
\midrule
\multicolumn{9}{c}{{\footnotesize{}AdLasso}}\tabularnewline
{\footnotesize{}True} & {\footnotesize{}0}&{\footnotesize{}712 (0.021)} & {\footnotesize{}0}&{\footnotesize{}751 (0.018)} & \multicolumn{2}{c}{} & \multicolumn{2}{c}{}\tabularnewline
{\footnotesize{}ARM} & {\footnotesize{}0}&{\footnotesize{}627 (0.020)} & {\footnotesize{}0}&{\footnotesize{}682 (0.016)} & {\footnotesize{}0}&{\footnotesize{}091 (0.006)} & {\footnotesize{}0}&{\footnotesize{}076 (0.005)}\tabularnewline
{\footnotesize{}BIC-p} & {\footnotesize{}0}&{\footnotesize{}716 (0.021)} & {\footnotesize{}0}&{\footnotesize{}754 (0.017)} & {\footnotesize{}0}&{\footnotesize{}030 (0.006)} & {\footnotesize{}0}&{\footnotesize{}026 (0.005)}\tabularnewline
\midrule
\multicolumn{9}{c}{{\footnotesize{}MCP}}\tabularnewline
{\footnotesize{}True} & {\footnotesize{}0}&{\footnotesize{}498 (0.015)} & {\footnotesize{}0}&{\footnotesize{}576 (0.012)} & \multicolumn{2}{c}{} & \multicolumn{2}{c}{}\tabularnewline
{\footnotesize{}ARM} & {\footnotesize{}0}&{\footnotesize{}433 (0.015)} & {\footnotesize{}0}&{\footnotesize{}523 (0.012)} & {\footnotesize{}0}&{\footnotesize{}067 (0.004)} & {\footnotesize{}0}&{\footnotesize{}056 (0.003)}\tabularnewline
{\footnotesize{}BIC-p} & {\footnotesize{}0}&{\footnotesize{}511 (0.015)} & {\footnotesize{}0}&{\footnotesize{}586 (0.012)} & {\footnotesize{}0}&{\footnotesize{}026 (0.005)} & {\footnotesize{}0}&{\footnotesize{}020 (0.004)}\tabularnewline
\midrule
\multicolumn{9}{c}{{\footnotesize{}SCAD}}\tabularnewline
{\footnotesize{}True} & {\footnotesize{}0}&{\footnotesize{}214 (0.006)} & {\footnotesize{}0}&{\footnotesize{}344 (0.005)} & \multicolumn{2}{c}{} & \multicolumn{2}{c}{}\tabularnewline
{\footnotesize{}ARM} & {\footnotesize{}0}&{\footnotesize{}183 (0.006)} & {\footnotesize{}0}&{\footnotesize{}312 (0.006)} & {\footnotesize{}0}&{\footnotesize{}032 (0.002)} & {\footnotesize{}0}&{\footnotesize{}033 (0.002)}\tabularnewline
{\footnotesize{}BIC-p} & {\footnotesize{}0}&{\footnotesize{}225 (0.007)} & {\footnotesize{}0}&{\footnotesize{}352 (0.006)} & {\footnotesize{}0}&{\footnotesize{}017 (0.004)} & {\footnotesize{}0}&{\footnotesize{}014 (0.003)}\tabularnewline
\bottomrule
\end{tabular}
\par\end{centering}{\footnotesize \par}

\end{table}

\begin{table}
\caption{Classification case (Example 4): $n=200$, $p=30$, the components
1--5 of $\boldsymbol{\beta}$ are 10.5, components 6--10
are 5.5, components 11--15 are 0.5 and the rests are zeros.
$\mathbf{x}_{i}$ for $i=1\protect\ddd n$ are from $X\sim N_{p}(0,\boldsymbol{\Sigma})$
with $\boldsymbol{\Sigma}=(0.4^{|j-k|})_{p\times p}$. Values are
averaged over 100 independent runs. The standard errors are shown
in parentheses. \label{tab:example-4}}
\begin{centering}
{\footnotesize{}}%
\begin{tabular}{lr@{\extracolsep{0pt}.}lr@{\extracolsep{0pt}.}lr@{\extracolsep{0pt}.}lr@{\extracolsep{0pt}.}l}
\toprule 
 & \multicolumn{2}{c}{{\footnotesize{}$F$}} & \multicolumn{2}{c}{{\footnotesize{}$G$}} & \multicolumn{2}{c}{{\footnotesize{}$d_{F}$}} & \multicolumn{2}{c}{{\footnotesize{}$d_{G}$}}\tabularnewline
\midrule 
\multicolumn{9}{c}{{\footnotesize{}Lasso}}\tabularnewline
{\footnotesize{}True} & {\footnotesize{}0}&{\footnotesize{}720 (0.005)} & {\footnotesize{}0}&{\footnotesize{}734 (0.005)} & \multicolumn{2}{c}{} & \multicolumn{2}{c}{}\tabularnewline
{\footnotesize{}ARM} & {\footnotesize{}0}&{\footnotesize{}493 (0.006)} & {\footnotesize{}0}&{\footnotesize{}572 (0.004)} & {\footnotesize{}0}&{\footnotesize{}227 (0.007)} & {\footnotesize{}0}&{\footnotesize{}163 (0.006)}\tabularnewline
{\footnotesize{}BIC-p} & {\footnotesize{}0}&{\footnotesize{}616 (0.006)} & {\footnotesize{}0}&{\footnotesize{}667 (0.004)} & {\footnotesize{}0}&{\footnotesize{}109 (0.005)} & {\footnotesize{}0}&{\footnotesize{}077 (0.005)}\tabularnewline
\midrule
\multicolumn{9}{c}{{\footnotesize{}AdLasso}}\tabularnewline
{\footnotesize{}True} & {\footnotesize{}0}&{\footnotesize{}794 (0.005)} & {\footnotesize{}0}&{\footnotesize{}800 (0.005)} & \multicolumn{2}{c}{} & \multicolumn{2}{c}{}\tabularnewline
{\footnotesize{}ARM} & {\footnotesize{}0}&{\footnotesize{}722 (0.006)} & {\footnotesize{}0}&{\footnotesize{}755 (0.005)} & {\footnotesize{}0}&{\footnotesize{}081 (0.006)} & {\footnotesize{}0}&{\footnotesize{}059 (0.005)}\tabularnewline
{\footnotesize{}BIC-p} & {\footnotesize{}0}&{\footnotesize{}876 (0.006)} & {\footnotesize{}0}&{\footnotesize{}883 (0.005)} & {\footnotesize{}0}&{\footnotesize{}096 (0.006)} & {\footnotesize{}0}&{\footnotesize{}094 (0.006)}\tabularnewline
\midrule
\multicolumn{9}{c}{{\footnotesize{}MCP}}\tabularnewline
{\footnotesize{}True} & {\footnotesize{}0}&{\footnotesize{}751 (0.005)} & {\footnotesize{}0}&{\footnotesize{}770 (0.005)} & \multicolumn{2}{c}{} & \multicolumn{2}{c}{}\tabularnewline
{\footnotesize{}ARM} & {\footnotesize{}0}&{\footnotesize{}793 (0.004)} & {\footnotesize{}0}&{\footnotesize{}813 (0.004)} & {\footnotesize{}0}&{\footnotesize{}063 (0.005)} & {\footnotesize{}0}&{\footnotesize{}056 (0.004)}\tabularnewline
{\footnotesize{}BIC-p} & {\footnotesize{}0}&{\footnotesize{}932 (0.005)} & {\footnotesize{}0}&{\footnotesize{}934 (0.005)} & {\footnotesize{}0}&{\footnotesize{}182 (0.006)} & {\footnotesize{}0}&{\footnotesize{}164 (0.005)}\tabularnewline
\midrule
\multicolumn{9}{c}{{\footnotesize{}SCAD}}\tabularnewline
{\footnotesize{}True} & {\footnotesize{}0}&{\footnotesize{}778 (0.006)} & {\footnotesize{}0}&{\footnotesize{}789 (0.006)} & \multicolumn{2}{c}{} & \multicolumn{2}{c}{}\tabularnewline
{\footnotesize{}ARM} & {\footnotesize{}0}&{\footnotesize{}755 (0.005)} & {\footnotesize{}0}&{\footnotesize{}781 (0.004)} & {\footnotesize{}0}&{\footnotesize{}064 (0.006)} & {\footnotesize{}0}&{\footnotesize{}055 (0.005)}\tabularnewline
{\footnotesize{}BIC-p} & {\footnotesize{}0}&{\footnotesize{}913 (0.006)} & {\footnotesize{}0}&{\footnotesize{}916 (0.005)} & {\footnotesize{}0}&{\footnotesize{}141 (0.007)} & {\footnotesize{}0}&{\footnotesize{}132 (0.006)}\tabularnewline
\bottomrule
\end{tabular}
\par\end{centering}{\footnotesize \par}

\end{table}

\begin{table}
\caption{Classification case (Example 5): $n=200$, $p=200$, the components
1--5 of $\beta$ are 10.5, the components 6--10
are 5.5, the components 11--15 are 0.5 and the rests are
zeros. $\mathbf{x}_{i}$ for $i=1\protect\ddd n$ are from $X\sim N_{p}(0,\boldsymbol{\Sigma})$.
The first 15 predictors $(X_{1},\ldots,X_{15})$ and the remaining
185 predictors $(X_{16},\ldots,X_{200})$ are independent. The correlation
between $X_{j}$ and $X_{k}$ in $(X_{1},\ldots,X_{15})$ is $0.4^{|j-k|}$.
The correlation between $X_{j}$ and $X_{k}$ in $(X_{16},\ldots,X_{200})$
is $0.4^{|j-k|}$. Values are averaged over 100 independent runs.
The standard errors are shown in parentheses. \label{tab:example-5}}
\begin{centering}
{\footnotesize{}}%
\begin{tabular}{lr@{\extracolsep{0pt}.}lr@{\extracolsep{0pt}.}lr@{\extracolsep{0pt}.}lr@{\extracolsep{0pt}.}l}
\toprule 
 & \multicolumn{2}{c}{{\footnotesize{}$F$}} & \multicolumn{2}{c}{{\footnotesize{}$G$}} & \multicolumn{2}{c}{{\footnotesize{}$d_{F}$}} & \multicolumn{2}{c}{{\footnotesize{}$d_{G}$}}\tabularnewline
\midrule 
\multicolumn{9}{c}{{\footnotesize{}Lasso}}\tabularnewline
{\footnotesize{}True} & {\footnotesize{}0}&{\footnotesize{}386 (0.006)} & {\footnotesize{}0}&{\footnotesize{}440 (0.005)} & \multicolumn{2}{c}{} & \multicolumn{2}{c}{}\tabularnewline
{\footnotesize{}ARM} & {\footnotesize{}0}&{\footnotesize{}223 (0.004)} & {\footnotesize{}0}&{\footnotesize{}348 (0.004)} & {\footnotesize{}0}&{\footnotesize{}163 (0.006)} & {\footnotesize{}0}&{\footnotesize{}093 (0.005)}\tabularnewline
{\footnotesize{}BIC-p} & {\footnotesize{}0}&{\footnotesize{}359 (0.006)} & {\footnotesize{}0}&{\footnotesize{}465 (0.005)} & {\footnotesize{}0}&{\footnotesize{}039 (0.004)} & {\footnotesize{}0}&{\footnotesize{}043 (0.003)}\tabularnewline
\midrule
\multicolumn{9}{c}{{\footnotesize{}AdLasso}}\tabularnewline
{\footnotesize{}True} & {\footnotesize{}0}&{\footnotesize{}726 (0.005)} & {\footnotesize{}0}&{\footnotesize{}735 (0.005)} & \multicolumn{2}{c}{} & \multicolumn{2}{c}{}\tabularnewline
{\footnotesize{}ARM} & {\footnotesize{}0}&{\footnotesize{}616 (0.008)} & {\footnotesize{}0}&{\footnotesize{}669 (0.006)} & {\footnotesize{}0}&{\footnotesize{}118 (0.007)} & {\footnotesize{}0}&{\footnotesize{}079 (0.005)}\tabularnewline
{\footnotesize{}BIC-p} & {\footnotesize{}0}&{\footnotesize{}859 (0.008)} & {\footnotesize{}0}&{\footnotesize{}865 (0.008)} & {\footnotesize{}0}&{\footnotesize{}137 (0.007)} & {\footnotesize{}0}&{\footnotesize{}133 (0.006)}\tabularnewline
\midrule
\multicolumn{9}{c}{{\footnotesize{}MCP}}\tabularnewline
{\footnotesize{}True} & {\footnotesize{}0}&{\footnotesize{}683 (0.008)} & {\footnotesize{}0}&{\footnotesize{}695 (0.008)} & \multicolumn{2}{c}{} & \multicolumn{2}{c}{}\tabularnewline
{\footnotesize{}ARM} & {\footnotesize{}0}&{\footnotesize{}639 (0.009)} & {\footnotesize{}0}&{\footnotesize{}687 (0.007)} & {\footnotesize{}0}&{\footnotesize{}079 (0.006)} & {\footnotesize{}0}&{\footnotesize{}063 (0.005)}\tabularnewline
{\footnotesize{}BIC-p} & {\footnotesize{}0}&{\footnotesize{}868 (0.008)} & {\footnotesize{}0}&{\footnotesize{}871 (0.008)} & {\footnotesize{}0}&{\footnotesize{}186 (0.006)} & {\footnotesize{}0}&{\footnotesize{}177 (0.006)}\tabularnewline
\midrule
\multicolumn{9}{c}{{\footnotesize{}SCAD}}\tabularnewline
{\footnotesize{}True} & {\footnotesize{}0}&{\footnotesize{}634 (0.008)} & {\footnotesize{}0}&{\footnotesize{}637 (0.008)} & \multicolumn{2}{c}{} & \multicolumn{2}{c}{}\tabularnewline
{\footnotesize{}ARM} & {\footnotesize{}0}&{\footnotesize{}506 (0.010)} & {\footnotesize{}0}&{\footnotesize{}580 (0.008)} & {\footnotesize{}0}&{\footnotesize{}131 (0.007)} & {\footnotesize{}0}&{\footnotesize{}072 (0.005)}\tabularnewline
{\footnotesize{}BIC-p} & {\footnotesize{}0}&{\footnotesize{}743 (0.009)} & {\footnotesize{}0}&{\footnotesize{}766 (0.008)} & {\footnotesize{}0}&{\footnotesize{}110 (0.006)} & {\footnotesize{}0}&{\footnotesize{}130 (0.006)}\tabularnewline
\bottomrule
\end{tabular}
\par\end{centering}{\footnotesize \par}

\end{table}

\subsection{Setting II: regression models}

For the regression case, the response $Y$ is generated from the following
model
\[
Y=X\boldsymbol{\beta}+\epsilon,
\]
where $\epsilon\sim N(0,\sigma^{2})$. The explanatory variables $X$
and the coefficient vector $\boldsymbol{\beta}$ are set under the
same settings as in the classification cases \ref{eg:fg-1}--\ref{eg:fg-5}.
To study how the estimation performances vary with the noise level
$\sigma^{2}$, we choose nine $\sigma$-values evenly spaced between
$0.01$ and $5$. 

\begin{figure}
\begin{centering}
\includegraphics[scale=0.8]{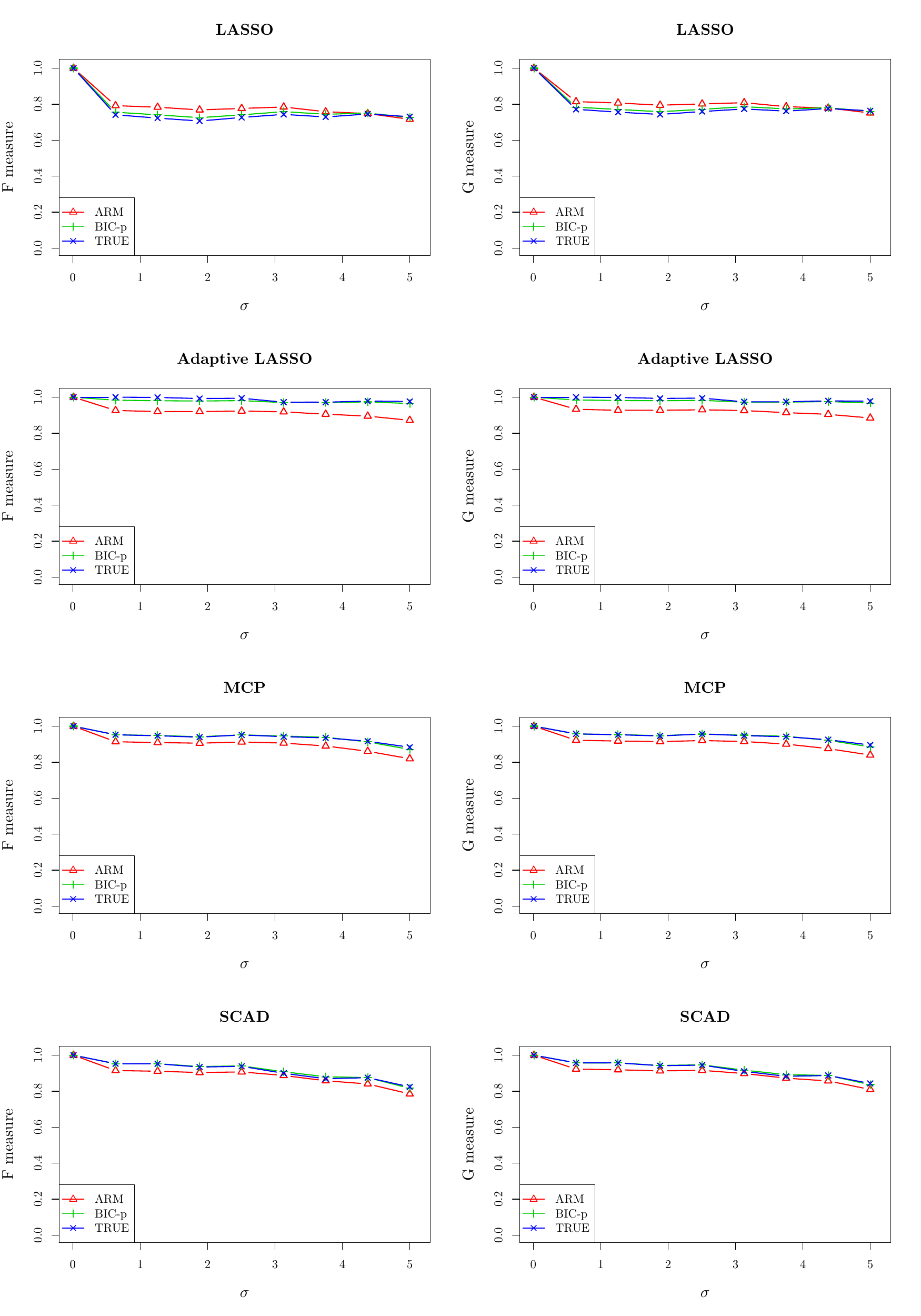}
\par\end{centering}
\caption{Regression case (Example 1): $n=200$, $p=8$, $\boldsymbol{\beta}=(3,1.5,2,0,0,0,0,0)^{\intercal}.$
$\mathbf{x}_{i}$ for $i=1\protect\ddd n$ are i.i.d. from $N_{p}(0,\boldsymbol{I})$.
The first column presents the results for $F$-measure, the second
column is for $G$-measure. \label{figure: regression-example-1}}
\end{figure}

\begin{figure}
\centering{}\includegraphics[scale=0.8]{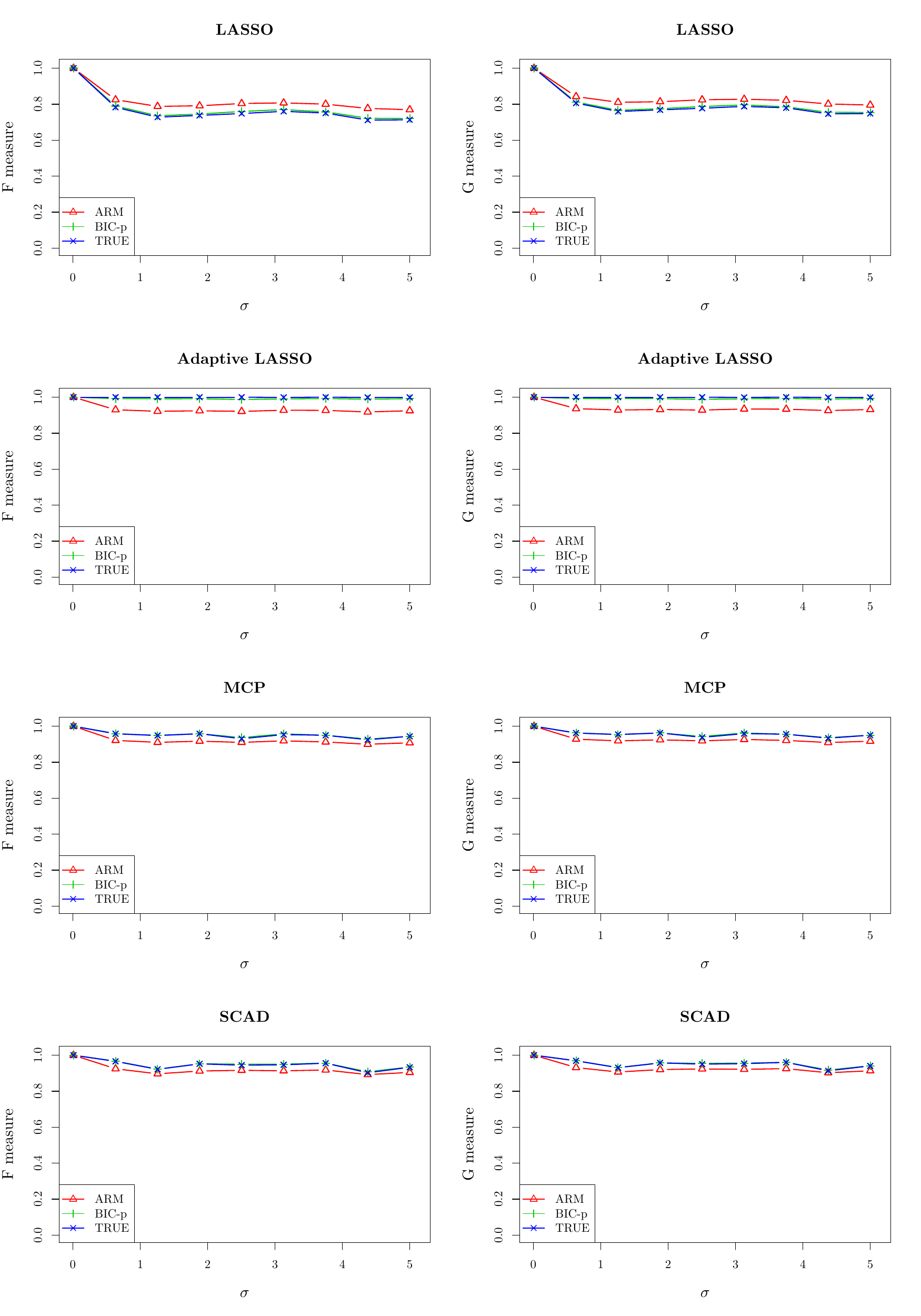}\caption{Regression case (Example 2): $n=1000$, $p=8$, $\boldsymbol{\beta}=(3,1.5,2,0,0,0,0,0)^{\intercal}.$
$\mathbf{x}_{i}$ for $i=1\protect\ddd n$ are i.i.d. from $N_{p}(0,\boldsymbol{I})$.
The first column presents the results for $F$-measure, the second
column is for $G$-measure.\label{figure: regression-example-2}}
\end{figure}

\begin{figure}
\centering{}\includegraphics[scale=0.8]{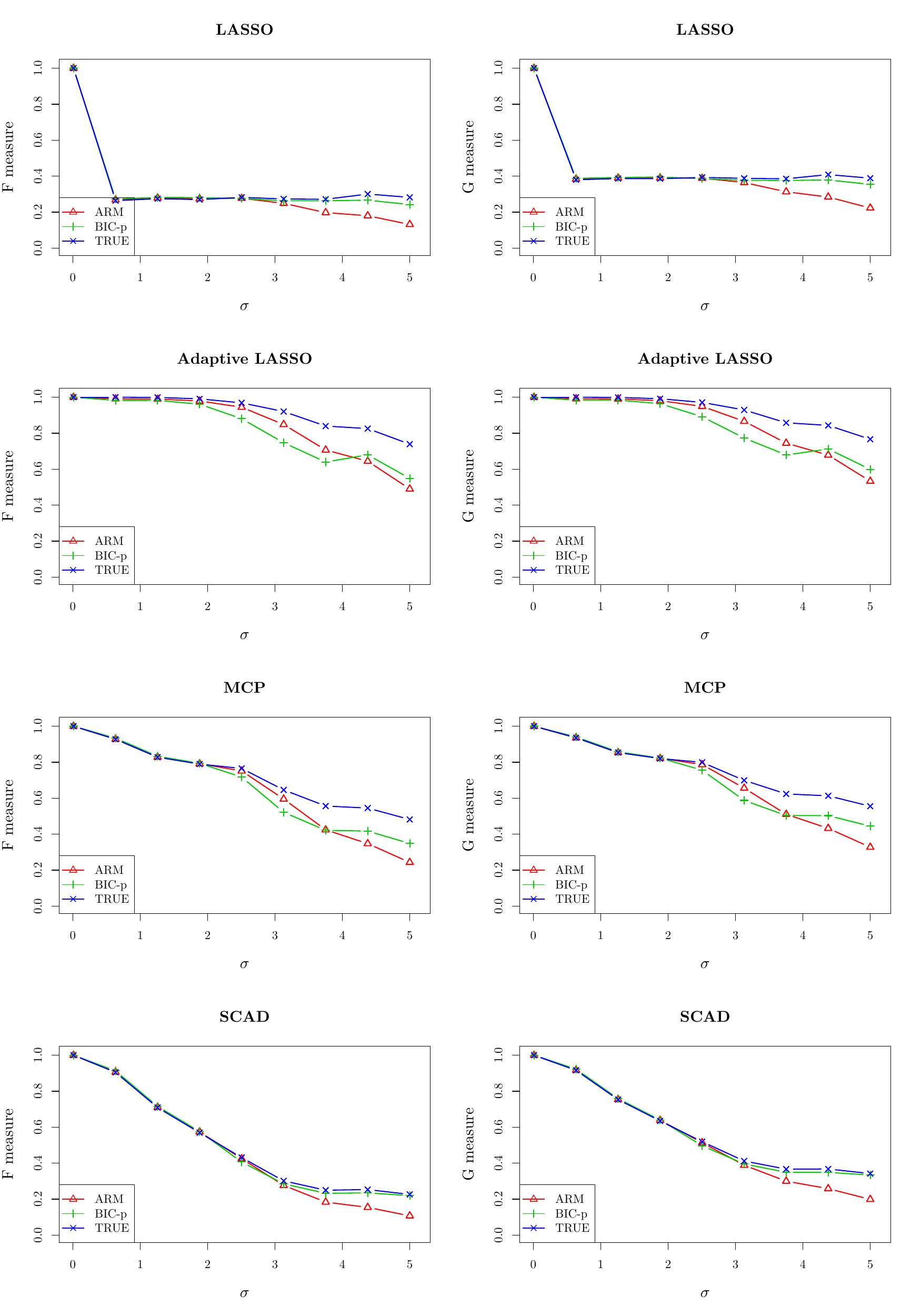}\caption{Regression case (Example 3): $n=200$, $p=2000$, $\boldsymbol{\beta}=(\beta_{1},\ldots,\beta_{p})^{\intercal}$,
where $(\beta_{1},\beta_{2},\beta_{3})=(3,1.5,2)$ and $(\beta_{4},\ldots,\beta_{2000})$
are zeros. $\mathbf{x}_{i}$ for $i=1\protect\ddd n$ are i.i.d. from
$N(0,\boldsymbol{I}_{p})$. The first column presents the results
for $F$-measure, the second column is for $G$-measure.\label{figure: regression-example-3}}
\end{figure}

\begin{figure}
\begin{centering}
\includegraphics[scale=0.8]{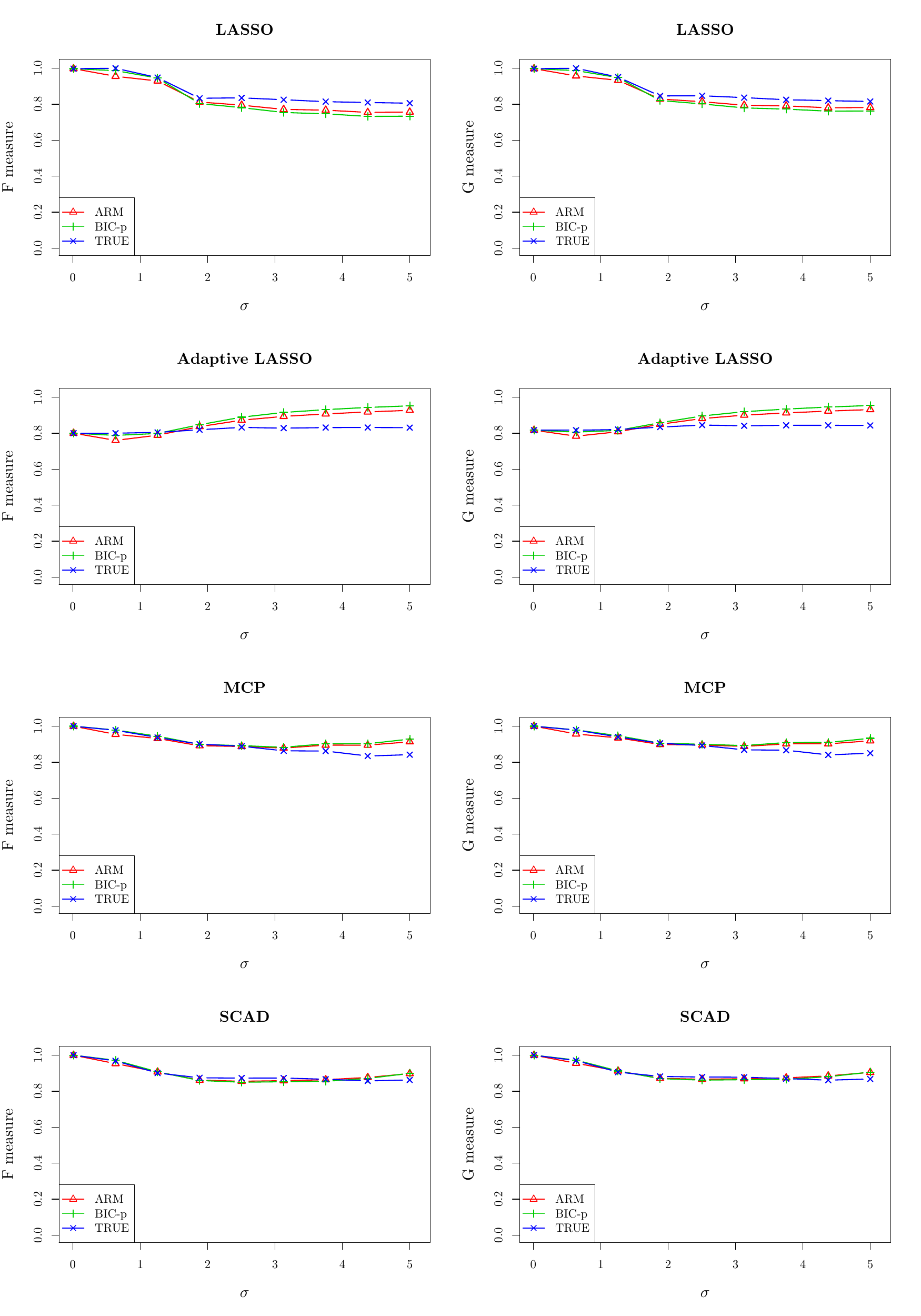}
\par\end{centering}
\caption{Regression case (Example 4): $n=200$, $p=30$, the components 1--5
of $\boldsymbol{\beta}$ are 10.5, components 6--10 are
5.5, components 11--15 are 0.5 and the rests are zeros.
$\mathbf{x}_{i}$ for $i=1\protect\ddd n$ are from $X\sim N_{p}(0,\boldsymbol{\Sigma})$
with $\boldsymbol{\Sigma}=(0.4^{|j-k|})_{p\times p}$. The first column
presents the results for $F$-measure, the second column is for $G$-measure.\label{figure: regression-example-4}}
\end{figure}

\begin{figure}
\begin{centering}
\includegraphics[scale=0.8]{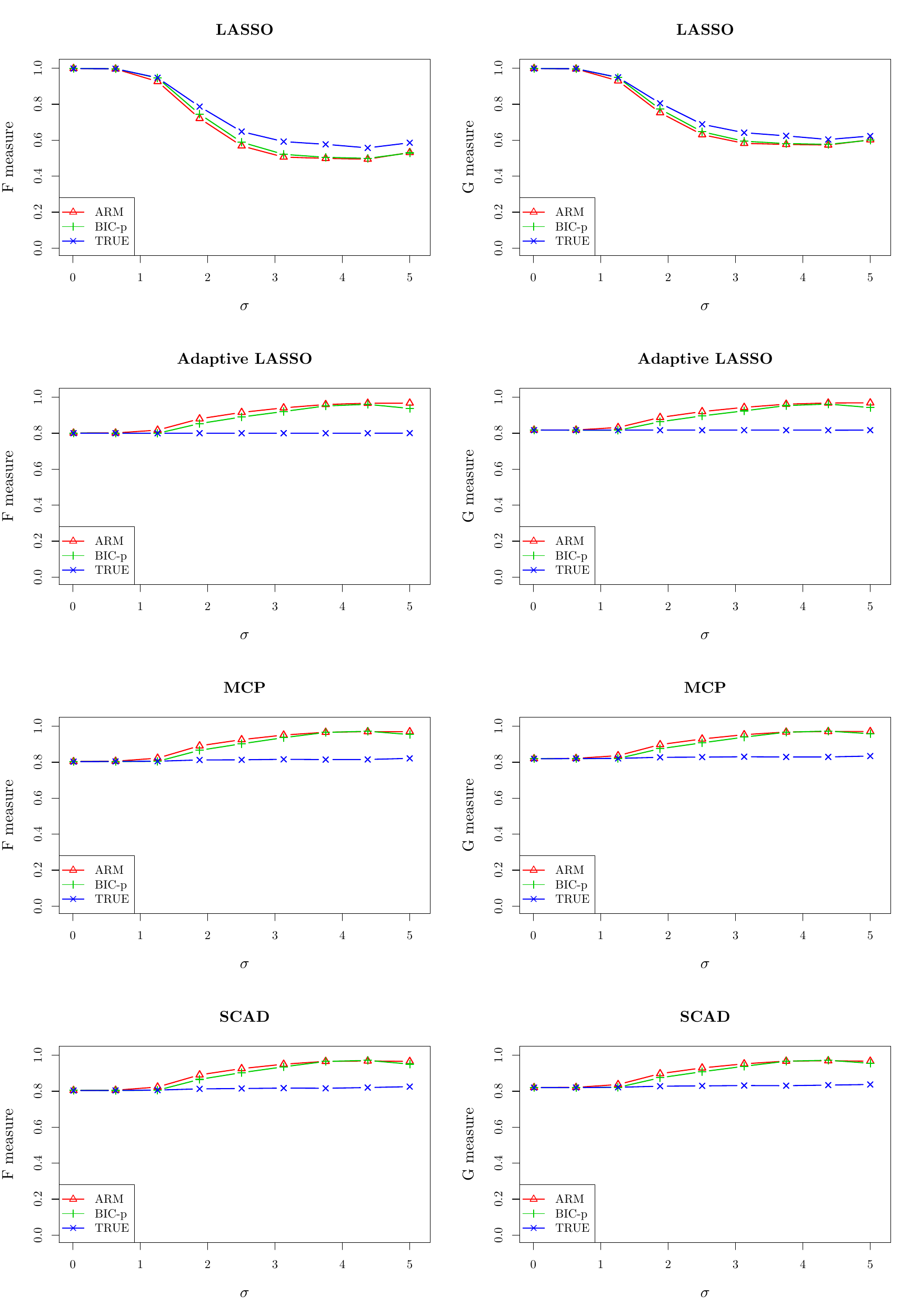}
\par\end{centering}
\caption{Regression case (Example 5): $n=200$, $p=200$, the components 1--5
of $\beta$ are 10.5, the components 6--10 are 5.5, the
components 11--15 are 0.5 and the rests are zeros. $\mathbf{x}_{i}$
for $i=1\protect\ddd n$ are from $X\sim N_{p}(0,\boldsymbol{\Sigma})$.
The first 15 predictors $(X_{1},\ldots,X_{15})$ and the remaining
185 predictors $(X_{16},\ldots,X_{200})$ are independent. The correlation
between $X_{j}$ and $X_{k}$ in $(X_{1},\ldots,X_{15})$ is $0.4^{|j-k|}$.
The correlation between $X_{j}$ and $X_{k}$ in $(X_{16},\ldots,X_{200})$
is $0.4^{|j-k|}$.  The first column presents the results for $F$-measure,
the second column is for $G$-measure.\label{figure: regression-example-5}}
\end{figure}

We compare $\widehat{F}(\mathcal{A}^{0})$ and $\widehat{G}(\mathcal{A}^{0})$
with the true $F(\az)$ and $G(\az)$ in Figures \ref{figure: regression-example-1}--\ref{figure: regression-example-5}. Overall, $\widehat{F}(\mathcal{A}^{0})$
and $\widehat{G}(\mathcal{A}^{0})$ using ARM and BIC-p weighting
can well reflect the trends of $F(\az)$ and $G(\az)$ in the sense
that, both the true curves and the estimated curves trend down as
$\sigma^{2}$ increases. And the estimation accuracy drops as $\sigma^{2}$
increases. The estimated $\widehat{F}(\mathcal{A}^{0})$ and $\widehat{G}(\mathcal{A}^{0})$
properly reflect the true performance of a given $\az$. For example,
in Figure \ref{figure: regression-example-3}--\ref{figure: regression-example-5},
we see that the performance of Lasso deteriorates significantly as
$\sigma^{2}$ increases, due to the fact that it tends to over-select
variables under higher noise levels. In contrast, adaptive Lasso,
MCP and SCAD have more robust performance against the high noise.
$\widehat{F}(\mathcal{A}^{0})$ and $\widehat{G}(\mathcal{A}^{0})$
can correctly reflect these aforementioned facts. From the results,
we find that MCP is the best performer with the highest true/estimated
$F$- and $G$-measures in Example 2-5, while adaptive Lasso is the
best performer in Example 1. 

By comparing Figures \ref{figure: regression-example-1} and \ref{figure: regression-example-2},
we see that the sample size influences the estimation performance:
large samples produce more accurate $\widehat{F}(\mathcal{A}^{0})$
and $\widehat{G}(\mathcal{A}^{0})$. Gains in the estimation accuracy
from increased sample sizes are due to the fact that more information
results in better assigned weights on the candidate models. 

In Figure 5, the over-estimation in SCAD and MCP, when $\sigma$ is
large, is due to highly weighted candidate models miss several small
coefficients variables, which is caused by the decaying coefficients
and worsened by correlation between the variables. While for Lasso,
when $\sigma$ is small, PAVI can find good candidate models to put
high weights on, thus the estimation is good; when $\sigma$ is larger,
the candidate models with high weights miss several true variables.
At the same time, Lasso chooses more redundant variables when $\sigma$
becomes larger. Therefore, the precision is under-estimated, so does
the $F$-measure.

\section{REAL DATA}

In this section, we apply PAVI to several model selection methods
using gene expression data for cancer-related biomarker identification.
The biomarker selection process is usually under high-dimensional,
small-sample, and high-noise setting with highly-correlated genes
involved \citep{golub1999,west2001predicting,ma2008penalized,ang2015supervised}.
As such, the sets of genes identified may be subject to substantial
changes due to small perturbations in the data \citep{baggerly2004reproducibility,BM2010,henry2012cancer,nanying,lim2015estimation,stodden2015reproducing}.
Here we use $\widehat{F}$ and $\widehat{G}$ to evaluate such selection
uncertainty. 

Our goal is to provide a serious and careful analysis of outcomes
of variable selection methods from multiple angles to understand the
key issues of interest. One may wonder if any strong statement can
be said because no one knows the truth. We hope our analysis provides
strong enough evidence that the estimated $F$ and $G$ values yield
valuable information.

\subsection{Data description}

We consider three well-studied benchmark cancer datasets: \texttt{Colon}
\citep{alon1999broad}, \texttt{Leukemia} \citep{golub1999} and \texttt{Prostate}
\citep{singh2002}. Table \ref{tab:Summary-of-Colon,} provides a
brief summary.

\begin{table}
\caption{Summary of \texttt{Colon}, \texttt{Leukemia}, \texttt{Prostate}. In
\texttt{Colon}, $y=1$ represents colon tumor samples and $y=0$ represents
normal colon tissue samples; In \texttt{Leukemia}, $y=1$ represents
acute myeloblastic leukemia samples and $y=0$ represents acute lymphoblastic
leukemia samples; In \texttt{Prostate}, $y=1$ represents tumor samples
and $y=0$ represents normal prostate samples. \label{tab:Summary-of-Colon,}}
\begin{centering}
{\footnotesize{}}%
\begin{tabular}{llllll}
\toprule 
\multirow{2}{*}{{\footnotesize{}Data}} & \multirow{2}{*}{{\footnotesize{}$n$}} & {\footnotesize{}$n_{1}$} & {\footnotesize{}$n_{2}$} & {\footnotesize{}$p$} & \multirow{2}{*}{{\footnotesize{}Data source}}\tabularnewline
 &  & {\footnotesize{}($y=1$)} & {\footnotesize{}($y=0$)} & {\footnotesize{}(number of genes)} & \tabularnewline
\midrule
\texttt{\footnotesize{}Colon} & {\footnotesize{}62} & {\footnotesize{}40} & {\footnotesize{}22} & {\footnotesize{}2000} & {\footnotesize{}\citet{alon1999broad}}\tabularnewline
\texttt{\footnotesize{}Leukemia} & {\footnotesize{}72} & {\footnotesize{}25} & {\footnotesize{}47} & {\footnotesize{}7129} & {\footnotesize{}\citet{golub1999}}\tabularnewline
\texttt{\footnotesize{}Prostate} & {\footnotesize{}102} & {\footnotesize{}52} & {\footnotesize{}50} & {\footnotesize{}12600} & {\footnotesize{}\citet{singh2002}}\tabularnewline
\bottomrule
\end{tabular}
\par\end{centering}{\footnotesize \par}
\end{table}

\subsection{Methods/models to be examined}

On these three datasets, we compare the variable selection performance
of four commonly used penalized regression methods: Lasso, adaptive
Lasso, MCP and SCAD. We first obtain the final model $\az$ for each
method (the tuning parameter $\lambda$ is selected using five-fold
cross-validation). Then we use PAVI to estimate $\widehat{F}(\mathcal{A}^{0})$
and $\widehat{G}(\mathcal{A}^{0})$ with two weightings, ARM and BIC-p.
The whole procedure is repeated 100 times to average out randomness
in the tuning parameter selection, and the averages of $\widehat{F}(\mathcal{A}^{0})$,
$\mathrm{sd}\big(\widehat{F}(\mathcal{A}^{0})\big)$ and $\widehat{G}(\mathcal{A}^{0})$,
$\mathrm{sd}\big(\widehat{G}(\mathcal{A}^{0})\big)$ are summarized
in Tables \ref{tab: realdata-Colon}, \ref{tab: realdata-Leukemia}
and \ref{tab: realdata-Prostate}. For comparison, we also include
several other models studied in the existing literature. Specifically,
we consider \citealp{31} (L10), \citealp{91} (Y10), \citealp{78}
(C11) and \citealp{116} (L11) for \texttt{Colon}, \citealp{31} (L10),
\citealp{91} (Y10), and \citealp{93} (J11; two kinds of models are provided via different importance criterion in this work, denoted by J11$^{1}$ and J11$^{2}$ hereafter respectively) for \texttt{Leukemia},
and \citealp{31} (L10) and \citealp{52} (S12) for \texttt{Prostate}. 

Y10, J11 and S12 used linear-based variable selection techniques without
initial variable screening. Specifically, Y10 used the probit regression
model; J11 used the linear kernel support vector classifier (SVC);
S12 used the linear discriminant analysis (LDA) technique with nearest
centroid classifier (NCC). In contrast, L10, C11 and L11 used nonparametric
variable selection techniques: L10 used SVM; C11 used the na\"ive Bayes
classifier (NBC) and SVM; L11 used the support vector machine (SVM).
In addition, we consider the Importance Screening method (ImpS) by
\citet{soilYeYangYang}, which uses a sparsity oriented importance
learning for variable screening. 

\subsection{Results}

The estimated $\widehat{F}$ and $\widehat{G}$ of each model on \texttt{Colon},
\texttt{Leukemia} and \texttt{Prostate} are reported in Tables \ref{tab: realdata-Colon},
\ref{tab: realdata-Leukemia} and \ref{tab: realdata-Prostate} respectively.
We find that ImpS achieves almost the largest estimated $\widehat{F}$
and $\widehat{G}$ on all three data sets. L10 has basically zero
$\widehat{F}$ and $\widehat{G}$ for \texttt{Colon} and \texttt{Prostate}. J11$^{1}$ and J11$^{2}$ has basically zero $\widehat{F}$ and $\widehat{G}$ for \texttt{Leukemia}. (These cases are bolded in Tables \ref{tab: realdata-Colon}, \ref{tab: realdata-Leukemia} and \ref{tab: realdata-Prostate}.) This suggests that, from a logistic regression modeling perspective,
they may have chosen ``wrong'' variables and they have very low recalls
or precisions. 

\begin{table}
\caption{Estimated $F$- and $G$-measures and standard deviations for \texttt{Colon}.
L10 has numerically zero $\widehat{F}$ and $\widehat{G}$ values
(bolded in the Table). \label{tab: realdata-Colon}}
\begin{centering}
{\footnotesize{}}%
\begin{tabular}{llllllllll}
\toprule 
 & \multicolumn{4}{c}{{\footnotesize{}ARM}} &  & \multicolumn{4}{c}{{\footnotesize{}BIC-p}}\tabularnewline
\cmidrule{2-5} \cmidrule{7-10} 
 & {\footnotesize{}$F$} & {\footnotesize{}$sd.F$} & {\footnotesize{}$G$} & {\footnotesize{}$sd.G$} &  & {\footnotesize{}$F$} & {\footnotesize{}$sd.F$} & {\footnotesize{}$G$} & {\footnotesize{}$sd.G$}\tabularnewline
{\footnotesize{}Lasso} & {\footnotesize{}0.147} & {\footnotesize{}0.024} & {\footnotesize{}0.280} & {\footnotesize{}0.022} &  & {\footnotesize{}0.205} & {\footnotesize{}0.066} & {\footnotesize{}0.332} & {\footnotesize{}0.058}\tabularnewline
\multirow{1}{*}{{\footnotesize{}AdLasso}} & {\footnotesize{}0.194} & {\footnotesize{}0.165} & {\footnotesize{}0.255} & {\footnotesize{}0.211} &  & {\footnotesize{}0.309} & {\footnotesize{}0.191} & {\footnotesize{}0.361} & {\footnotesize{}0.209}\tabularnewline
\multirow{1}{*}{{\footnotesize{}MCP}} & {\footnotesize{}0.349} & {\footnotesize{}0.045} & {\footnotesize{}0.459} & {\footnotesize{}0.035} &  & {\footnotesize{}0.460} & {\footnotesize{}0.130} & {\footnotesize{}0.544} & {\footnotesize{}0.093}\tabularnewline
\multirow{1}{*}{{\footnotesize{}SCAD}} & {\footnotesize{}0.149} & {\footnotesize{}0.032} & {\footnotesize{}0.274} & {\footnotesize{}0.039} &  & {\footnotesize{}0.211} & {\footnotesize{}0.074} & {\footnotesize{}0.331} & {\footnotesize{}0.071}\tabularnewline
\multirow{1}{*}{{\footnotesize{}ImpS}} & {\footnotesize{}0.524} & {\footnotesize{}0.081} & {\footnotesize{}0.596} & {\footnotesize{}0.065} &  & {\footnotesize{}0.656} & {\footnotesize{}0.176} & {\footnotesize{}0.698} & {\footnotesize{}0.118}\tabularnewline
\multirow{1}{*}{{\footnotesize{}L11}} & {\footnotesize{}0.111} & {\footnotesize{}0.110} & {\footnotesize{}0.175} & {\footnotesize{}0.175} &  & {\footnotesize{}0.112} & {\footnotesize{}0.105} & {\footnotesize{}0.157} & {\footnotesize{}0.151}\tabularnewline
\multirow{1}{*}{{\footnotesize{}Y10}} & {\footnotesize{}0.103} & {\footnotesize{}0.017} & {\footnotesize{}0.233} & {\footnotesize{}0.018} &  & {\footnotesize{}0.146} & {\footnotesize{}0.048} & {\footnotesize{}0.276} & {\footnotesize{}0.047}\tabularnewline
\multirow{1}{*}{{\footnotesize{}C11}} & {\footnotesize{}0.184} & {\footnotesize{}0.020} & {\footnotesize{}0.317} & {\footnotesize{}0.022} &  & {\footnotesize{}0.223} & {\footnotesize{}0.076} & {\footnotesize{}0.333} & {\footnotesize{}0.082}\tabularnewline
\multirow{1}{*}{{\footnotesize{}L10}} & \textbf{\textcolor{black}{\footnotesize{}0.000}} & \textbf{\textcolor{black}{\footnotesize{}0.000}} & \textbf{\textcolor{black}{\footnotesize{}0.000}} & \textbf{\textcolor{black}{\footnotesize{}0.000}} &  & \textbf{\textcolor{black}{\footnotesize{}0.000}} & \textbf{\textcolor{black}{\footnotesize{}0.000}} & \textbf{\textcolor{black}{\footnotesize{}0.000}} & \textbf{\textcolor{black}{\footnotesize{}0.000}}\tabularnewline
\bottomrule
\end{tabular}
\par\end{centering}{\footnotesize \par}

\end{table}

\begin{table}
\caption{Estimated $F$- and $G$-measures and standard deviations for \texttt{Leukemia}.
J11$^{1}$ and J11$^{2}$  have numerically zero $\widehat{F}$ and $\widehat{G}$
values (bolded in the Table).\label{tab: realdata-Leukemia}}
\begin{centering}
{\footnotesize{}}%
\begin{tabular}{llllllllll}
\toprule 
 & \multicolumn{4}{c}{{\footnotesize{}ARM}} &  & \multicolumn{4}{c}{{\footnotesize{}BIC-p}}\tabularnewline
\cmidrule{2-5} \cmidrule{7-10} 
 & {\footnotesize{}$F$} & {\footnotesize{}$sd.F$} & {\footnotesize{}$G$} & {\footnotesize{}$sd.G$} &  & {\footnotesize{}$F$} & {\footnotesize{}$sd.F$} & {\footnotesize{}$G$} & {\footnotesize{}$sd.G$}\tabularnewline
{\footnotesize{}Lasso} & {\footnotesize{}0.083} & {\footnotesize{}0.025} & {\footnotesize{}0.206} & {\footnotesize{}0.026} &  & {\footnotesize{}0.079} & {\footnotesize{}0.012} & {\footnotesize{}0.203} & {\footnotesize{}0.014}\tabularnewline
\multirow{1}{*}{{\footnotesize{}AdLasso}} & {\footnotesize{}0.323} & {\footnotesize{}0.044} & {\footnotesize{}0.432} & {\footnotesize{}0.031} &  & {\footnotesize{}0.322} & {\footnotesize{}0.039} & {\footnotesize{}0.434} & {\footnotesize{}0.033}\tabularnewline
\multirow{1}{*}{{\footnotesize{}MCP}} & {\footnotesize{}0.168} & {\footnotesize{}0.170} & {\footnotesize{}0.221} & {\footnotesize{}0.210} &  & {\footnotesize{}0.061} & {\footnotesize{}0.089} & {\footnotesize{}0.078} & {\footnotesize{}0.108}\tabularnewline
\multirow{1}{*}{{\footnotesize{}SCAD}} & {\footnotesize{}0.094} & {\footnotesize{}0.028} & {\footnotesize{}0.220} & {\footnotesize{}0.028} &  & {\footnotesize{}0.090} & {\footnotesize{}0.013} & {\footnotesize{}0.216} & {\footnotesize{}0.015}\tabularnewline
\multirow{1}{*}{{\footnotesize{}ImpS}} & {\footnotesize{}0.525} & {\footnotesize{}0.065} & {\footnotesize{}0.591} & {\footnotesize{}0.042} &  & {\footnotesize{}0.573} & {\footnotesize{}0.129} & {\footnotesize{}0.636} & {\footnotesize{}0.102}\tabularnewline
\multirow{1}{*}{{\footnotesize{}$\mathrm{J11}^{1}$}} & \textbf{\textcolor{black}{\footnotesize{}0.000}} & \textbf{\textcolor{black}{\footnotesize{}0.000}} & \textbf{\textcolor{black}{\footnotesize{}0.000}} & \textbf{\textcolor{black}{\footnotesize{}0.000}} &  & \textbf{\textcolor{black}{\footnotesize{}0.000}} & \textbf{\textcolor{black}{\footnotesize{}0.000}} & \textbf{\textcolor{black}{\footnotesize{}0.000}} & \textbf{\textcolor{black}{\footnotesize{}0.000}}\tabularnewline
\multirow{1}{*}{{\footnotesize{}$\mathrm{J11}^{2}$}} & \textbf{\textcolor{black}{\footnotesize{}0.000}} & \textbf{\textcolor{black}{\footnotesize{}0.000}} & \textbf{\textcolor{black}{\footnotesize{}0.000}} & \textbf{\textcolor{black}{\footnotesize{}0.000}} &  & \textbf{\textcolor{black}{\footnotesize{}0.000}} & \textbf{\textcolor{black}{\footnotesize{}0.000}} & \textbf{\textcolor{black}{\footnotesize{}0.000}} & \textbf{\textcolor{black}{\footnotesize{}0.000}}\tabularnewline
\multirow{1}{*}{{\footnotesize{}Y10}} & {\footnotesize{}0.108} & {\footnotesize{}0.014} & {\footnotesize{}0.236} & {\footnotesize{}0.009} &  & {\footnotesize{}0.105} & {\footnotesize{}0.002} & {\footnotesize{}0.233} & {\footnotesize{}0.012}\tabularnewline
\multirow{1}{*}{{\footnotesize{}L10}} & {\footnotesize{}0.212} & {\footnotesize{}0.180} & {\footnotesize{}0.265} & {\footnotesize{}0.224} &  & {\footnotesize{}0.336} & {\footnotesize{}0.089} & {\footnotesize{}0.419} & {\footnotesize{}0.110}\tabularnewline
\bottomrule
\end{tabular}
\par\end{centering}{\footnotesize \par}
\end{table}

\begin{table}
\caption{Estimated $F$- and $G$-measures and standard deviations for \texttt{Prostate}.
L10 has numerically zero $\widehat{F}$ and $\widehat{G}$ values
(bolded in the Table). \label{tab: realdata-Prostate}}
\begin{centering}
{\footnotesize{}}%
\begin{tabular}{llllllllll}
\toprule 
 & \multicolumn{4}{c}{{\footnotesize{}ARM}} &  & \multicolumn{4}{c}{{\footnotesize{}BIC-p}}\tabularnewline
\cmidrule{2-5} \cmidrule{7-10} 
 & {\footnotesize{}$F$} & {\footnotesize{}$sd.F$} & {\footnotesize{}$G$} & {\footnotesize{}$sd.G$} &  & {\footnotesize{}$F$} & {\footnotesize{}$sd.F$} & {\footnotesize{}$G$} & {\footnotesize{}$sd.G$}\tabularnewline
{\footnotesize{}Lasso} & {\footnotesize{}0.064} & {\footnotesize{}0.004} & {\footnotesize{}0.181} & {\footnotesize{}0.005} &  & {\footnotesize{}0.064} & {\footnotesize{}0.003} & {\footnotesize{}0.181} & {\footnotesize{}0.004}\tabularnewline
\multirow{1}{*}{{\footnotesize{}AdLasso}} & {\footnotesize{}0.190} & {\footnotesize{}0.011} & {\footnotesize{}0.323} & {\footnotesize{}0.009} &  & {\footnotesize{}0.189} & {\footnotesize{}0.008} & {\footnotesize{}0.323} & {\footnotesize{}0.007}\tabularnewline
\multirow{1}{*}{{\footnotesize{}MCP}} & {\footnotesize{}0.018} & {\footnotesize{}0.019} & {\footnotesize{}0.027} & {\footnotesize{}0.022} &  & {\footnotesize{}0.018} & {\footnotesize{}0.012} & {\footnotesize{}0.027} & {\footnotesize{}0.014}\tabularnewline
\multirow{1}{*}{{\footnotesize{}SCAD}} & {\footnotesize{}0.097} & {\footnotesize{}0.006} & {\footnotesize{}0.225} & {\footnotesize{}0.007} &  & {\footnotesize{}0.096} & {\footnotesize{}0.005} & {\footnotesize{}0.225} & {\footnotesize{}0.005}\tabularnewline
\multirow{1}{*}{{\footnotesize{}ImpS}} & {\footnotesize{}0.333} & {\footnotesize{}0.011} & {\footnotesize{}0.447} & {\footnotesize{}0.008} &  & {\footnotesize{}0.333} & {\footnotesize{}0.012} & {\footnotesize{}0.447} & {\footnotesize{}0.009}\tabularnewline
\multirow{1}{*}{{\footnotesize{}S12}} & {\footnotesize{}0.395} & {\footnotesize{}0.037} & {\footnotesize{}0.494} & {\footnotesize{}0.047} &  & {\footnotesize{}0.400} & {\footnotesize{}0.003} & {\footnotesize{}0.500} & {\footnotesize{}0.007}\tabularnewline
\multirow{1}{*}{{\footnotesize{}L10}} & \textbf{\textcolor{black}{\footnotesize{}0.000}} & \textbf{\textcolor{black}{\footnotesize{}0.000}} & \textbf{\textcolor{black}{\footnotesize{}0.000}} & \textbf{\textcolor{black}{\footnotesize{}0.000}} &  & \textbf{\textcolor{black}{\footnotesize{}0.000}} & \textbf{\textcolor{black}{\footnotesize{}0.000}} & \textbf{\textcolor{black}{\footnotesize{}0.000}} & \textbf{\textcolor{black}{\footnotesize{}0.000}}\tabularnewline
\bottomrule
\end{tabular}
\par\end{centering}{\footnotesize \par}

\end{table}

\subsection{Are the zero $\widehat{F}$ and $\widehat{G}$ values too harsh for
the methods?}

It is striking that the $\widehat{F}$ and $\widehat{G}$ for some
selections are numerically zero, which seems rather extreme. Does
this mean those models are truly poor or rather our performance assessment
methodology fails? We would like to examine the matter from three
perspectives.

\subsubsection{First perspective: the labels of the selected genes}

First, let us examine the labels of the selected genes. We obtain
the selected genes in the literature. And we use five-fold cross-validation
in penalization parameter tuning to obtain selected genes for the
penalized regression models. In Tables \ref{tab: Gene ID-Colon},
\ref{tab: Gene ID-Leukemia} and \ref{tab: Gene ID-Prostate}, the
results show that the genes selected by L10 (\texttt{Colon} and \texttt{Prostate}), J11$^{1}$ and J11$^{2}$ (\texttt{Leukemia}) are mostly not
supported by other models. More specifically, the choices of variables by L10, J11$^{1}$ and J11$^{2}$ in those cases respectively
share zero, one or at most two genes with the other methods. (These cases are underlined in Tables \ref{tab: Gene ID-Colon}, \ref{tab: Gene ID-Leukemia} and \ref{tab: Gene ID-Prostate}.)

\begin{table}
\caption{Labels of selected genes for \texttt{Colon}. \label{tab: Gene ID-Colon}}
\begin{centering}
{\footnotesize{}}%
\begin{tabular}{ll}
\toprule 
 & {\footnotesize{}Labels of selected genes}\tabularnewline
\cmidrule{2-2} 
{\footnotesize{}Lasso} & \texttt{\footnotesize{}\{66, 249, 377, 493, 765, 1325, 1346, 1423, 1582,
1644, 1772, 1870\}}\tabularnewline
{\footnotesize{}AdLasso} & \texttt{\footnotesize{}\{249, 377, 765, 1582, 1772, 1870\}}\tabularnewline
{\footnotesize{}MCP} & \texttt{\footnotesize{}\{249, 377, 1644, 1772, 1870\}}\tabularnewline
\multirow{1}{*}{{\footnotesize{}SCAD}} & \texttt{\footnotesize{}\{377, 617, 765, 1024, 1325, 1346, 1482, 1504,
1582, 1644, 1772, 1870\}}\tabularnewline
{\footnotesize{}ImpS} & \texttt{\footnotesize{}\{249, 1772\}}\tabularnewline
{\footnotesize{}L11} & \texttt{\footnotesize{}\{249, 286, 765, 1058, 1485, 1671, 1771, 1836\}}\tabularnewline
\multirow{2}{*}{{\footnotesize{}Y10}} & \texttt{\footnotesize{}\{14, 161, 249, 377, 492, 493, 576, 792, 822,
1042, 1210, }\tabularnewline
 & \texttt{\footnotesize{}1346, 1400, 1423, 1549, 1635, 1772, 1843, 1924\}}\tabularnewline
{\footnotesize{}C11} & \texttt{\footnotesize{}\{249, 399, 513, 515, 780, 1042, 1325, 1582,
1771, 1772\}}\tabularnewline
{\footnotesize{}L10} & \texttt{\textbf{\textit{\emph{\footnotesize{}\uline{\{732, 994, 1473,
1763, 1794, 1843\}}}}}}\tabularnewline
\bottomrule
\end{tabular}
\par\end{centering}{\footnotesize \par}

\end{table}

\begin{table}
\caption{Labels of selected genes for \texttt{Leukemia}. \label{tab: Gene ID-Leukemia}}
\begin{centering}
{\footnotesize{}}%
\begin{tabular}{ll}
\toprule 
 & {\footnotesize{}Labels of selected genes}\tabularnewline
\cmidrule{2-2} 
\multirow{3}{*}{{\footnotesize{}Lasso}} & \texttt{\footnotesize{}\{804, 1239, 1674, 1745, 1779, 1796, 1834, 1882,
1928, 1933, }\tabularnewline
 & \texttt{\footnotesize{}1941, 2121, 2288, 3847, 4196, 4328, 4847, 4951,
4973, 5002, }\tabularnewline
 & \texttt{\footnotesize{}5107, 5335, 5766, 6055, 6169, 6539, 6855\}}\tabularnewline
{\footnotesize{}AdLasso} & \texttt{\footnotesize{}\{1779, 1834, 4328, 4847, 4951\}}\tabularnewline
{\footnotesize{}MCP} & \texttt{\footnotesize{}\{804, 1941, 3837, 4714, 4847, 4951, 6539\}}\tabularnewline
\multirow{2}{*}{{\footnotesize{}SCAD}} & \texttt{\footnotesize{}\{804, 1674, 1745, 1779, 1834, 1882, 1928, 1941,
2288, 3847, 4196, }\tabularnewline
 & \texttt{\footnotesize{}4328, 4847, 4951, 4973, 5002, 5766, 5772, 6169,
6225, 6281, 6539, 6855\}}\tabularnewline
{\footnotesize{}ImpS} & \texttt{\footnotesize{}\{1239, 4847, 4951\}}\tabularnewline
{\footnotesize{}$\mathrm{J11}^{1}$} & \texttt{\textcolor{black}{\footnotesize{}\uline{\{1376, 1394, 1674,
1882, 2186, 2402, 6200, 6201, 6803\}}}}\tabularnewline
{\footnotesize{}$\mathrm{J11}^{2}$} & \texttt{\textit{\textcolor{black}{\emph{\footnotesize{}\uline{\{1394,
1674, 1882, 2186, 5976, 6200, 6201, 6806\}}}}}}\tabularnewline
\multirow{2}{*}{{\footnotesize{}Y10}} & \texttt{\footnotesize{}\{760, 804, 1745, 1829, 1834, 1882, 2354, 3320,
4052, }\tabularnewline
 & \texttt{\footnotesize{}4211, 4377, 4535, 4847, 5039, 6041, 6218, 6376,
6540\}}\tabularnewline
{\footnotesize{}L10} & \texttt{\footnotesize{}\{220, 1086, 1834, 2020\}}\tabularnewline
\bottomrule
\end{tabular}
\par\end{centering}{\footnotesize \par}
\end{table}

\begin{table}
\caption{Labels of selected genes for \texttt{Prostate}. \label{tab: Gene ID-Prostate}}
\begin{centering}
{\footnotesize{}}%
\begin{tabular}{ll}
\toprule 
 & {\footnotesize{}Labels of selected genes}\tabularnewline
\cmidrule{2-2} 
\multirow{3}{*}{{\footnotesize{}Lasso}} & \texttt{\footnotesize{}\{1107, 3617, 4282, 4438, 4525, 4636, 5661, 5838,
5890, 6145, 6185, }\tabularnewline
 & \texttt{\footnotesize{}6838, 7375, 7428, 7539, 7623, 7915, 8123, 8965,
9034, 9093, 9816,}\tabularnewline
 & \texttt{\footnotesize{}9850, 10234, 10537, 10956, 11858, 11871, 12153,
12462\}}\tabularnewline
{\footnotesize{}AdLasso} & \texttt{\footnotesize{}\{5661, 5890, 6185, 7539, 7623, 8965, 9034, 9093,
10234, 11858\}}\tabularnewline
{\footnotesize{}MCP} & \texttt{\footnotesize{}\{7623, 7924, 8965, 9034, 9816, 10234, 11858\}}\tabularnewline
\multirow{2}{*}{{\footnotesize{}SCAD}} & \texttt{\footnotesize{}\{1107, 3540, 4636, 5661, 5838, 5890, 6185, 7623,
8603, 8965, 9034,}\tabularnewline
 & \texttt{\footnotesize{}9093, 9816, 10234, 10956, 11858, 11871, 12153\}}\tabularnewline
{\footnotesize{}ImpS} & \texttt{\footnotesize{}\{8965, 9034, 10234, 11858\}}\tabularnewline
{\footnotesize{}S12} & \texttt{\footnotesize{}\{4377, 6185, 6390, 6915\}}\tabularnewline
{\footnotesize{}L10} & \texttt{\textit{\textcolor{black}{\emph{\footnotesize{}\uline{\{4743,
6096, 8475, 9575, 9927, 12331\}}}}}}\tabularnewline
\bottomrule
\end{tabular}
\par\end{centering}{\footnotesize \par}

\end{table}

\subsubsection{Second perspective: predictive accuracy}

Secondly, we would like to examine the issue from a predictive accuracy
perspective. We randomly split the dataset into 4/5 observations for
training and 1/5 observations for testing. We fit the SVM models with
those selected genes on the training data using\textbf{ kernlab} \citep{zeileis2004kernlab}
and evaluate the predictive accuracy on the testing data. The whole
procedure is repeated 100 times and the averaged classification accuracy
and ``standard errors'' (w.r.t. the permutations) are recorded in
Table \ref{tab: Logistic}. Alternatively, we may
consider the parametric models. We fit the logistic regression with
the genes selected (in Table \ref{tab: Logistic}). We find that L10, J11$^{1}$ and J11$^{2}$ have worse predictive
accuracy (bolded in Table \ref{tab: Logistic}) compared with the simpler model by ImpS, which adds evidence
to support the validity of their low $\widehat{F}$ and $\widehat{G}$
values.

\begin{table}
\caption{Comparisons of classification accuracy on \texttt{Colon}, \texttt{Leukemia},
and \texttt{Prostate} using logistic regression and SVM, respectively. \label{tab: Logistic}}
\begin{centering}
{\footnotesize{}}%
\begin{tabular}{lllllllc}
\toprule 
\multicolumn{8}{c}{\textbf{\footnotesize{}Logistic Model}}\tabularnewline
\midrule
\multicolumn{2}{c}{\texttt{\footnotesize{}Colon}} &  & \multicolumn{2}{c}{\texttt{\footnotesize{}Leukemia}} &  & \multicolumn{2}{c}{\texttt{\footnotesize{}Prostate}}\tabularnewline
\cmidrule{1-2} \cmidrule{4-5} \cmidrule{7-8} 
{\footnotesize{}ImpS} & {\footnotesize{}86.3 (0.8)} &  & {\footnotesize{}ImpS} & {\footnotesize{}97.1 (0.3)} &  & {\footnotesize{}ImpS} & {\footnotesize{}94.0 (0.4)}\tabularnewline
{\footnotesize{}Lasso} & {\footnotesize{}80.0 (1.0)} &  & {\footnotesize{}Lasso} & {\footnotesize{}99.8 (0.1)} &  & {\footnotesize{}Lasso} & {\footnotesize{}97.0 (0.4)}\tabularnewline
{\footnotesize{}AdLasso} & {\footnotesize{}85.5 (0.8)} &  & {\footnotesize{}AdLasso} & {\footnotesize{}93.9 (0.5)} &  & {\footnotesize{}AdLasso} & {\footnotesize{}99.8 (0.1)}\tabularnewline
{\footnotesize{}MCP} & {\footnotesize{}85.1 (0.8)} &  & {\footnotesize{}MCP} & {\footnotesize{}99.5 (0.1)} &  & {\footnotesize{}MCP} & {\footnotesize{}98.7 (0.2)}\tabularnewline
{\footnotesize{}SCAD} & {\footnotesize{}84.3 (0.8)} &  & {\footnotesize{}SCAD} & {\footnotesize{}97.9 (0.3)} &  & {\footnotesize{}SCAD} & {\footnotesize{}97.1 (0.2)}\tabularnewline
{\footnotesize{}L11} & {\footnotesize{}80.4 (0.8)} &  & {\footnotesize{}$\mathrm{J11}^{1}$} & \textbf{\textcolor{black}{\footnotesize{}89.4 (0.8)}} &  & {\footnotesize{}S12} & {\footnotesize{}96.5 (0.4)}\tabularnewline
{\footnotesize{}Y10} & {\footnotesize{}90.9 (0.9)} &  & {\footnotesize{}$\mathrm{J11}^{2}$} & \textbf{\textcolor{black}{\footnotesize{}89.8 (0.7)}} &  & {\footnotesize{}L10} & \textbf{\textcolor{black}{\footnotesize{}59.0 (0.8)}}\tabularnewline
{\footnotesize{}C11} & {\footnotesize{}79.6 (1.0)} &  & {\footnotesize{}Y10} & {\footnotesize{}91.2 (0.7)} &  &  & \tabularnewline
{\footnotesize{}L10} & \textbf{\textcolor{black}{\footnotesize{}83.0 (0.9)}} &  & {\footnotesize{}L10} & {\footnotesize{}95.5 (0.4)} &  &  & \tabularnewline
\midrule
\multicolumn{8}{c}{\textbf{\footnotesize{}SVM Model}}\tabularnewline
\midrule
\multicolumn{2}{c}{\texttt{\footnotesize{}Colon}} &  & \multicolumn{2}{c}{\texttt{\footnotesize{}Leukemia}} &  & \multicolumn{2}{c}{\texttt{\footnotesize{}Prostate}}\tabularnewline
\cmidrule{1-2} \cmidrule{4-5} \cmidrule{7-8} 
{\footnotesize{}ImpS} & {\footnotesize{}84.0 (0.9)} &  & {\footnotesize{}ImpS} & {\footnotesize{}97.6 (0.3)} &  & {\footnotesize{}ImpS} & {\footnotesize{}95.3 (0.4)}\tabularnewline
{\footnotesize{}Lasso} & {\footnotesize{}75.8 (0.9)} &  & {\footnotesize{}Lasso} & {\footnotesize{}99.1 (0.2)} &  & {\footnotesize{}Lasso} & {\footnotesize{}96.3 (0.4)}\tabularnewline
{\footnotesize{}AdLasso} & {\footnotesize{}79.0 (0.9)} &  & {\footnotesize{}AdLasso} & {\footnotesize{}95.8 (0.4)} &  & {\footnotesize{}AdLasso} & {\footnotesize{}96.6 (0.3)}\tabularnewline
{\footnotesize{}MCP} & {\footnotesize{}83.1 (1.1)} &  & {\footnotesize{}MCP} & {\footnotesize{}99.0 (0.2)} &  & {\footnotesize{}MCP} & {\footnotesize{}97.1 (0.3)}\tabularnewline
{\footnotesize{}SCAD} & {\footnotesize{}86.0 (0.9)} &  & {\footnotesize{}SCAD} & {\footnotesize{}99.1 (0.2)} &  & {\footnotesize{}SCAD} & {\footnotesize{}96.4 (0.3)}\tabularnewline
{\footnotesize{}L11} & {\footnotesize{}79.0 (1.1)} &  & {\footnotesize{}$\mathrm{J11}^{1}$} & \textbf{\textcolor{black}{\footnotesize{}88.6 (0.8)}} &  & {\footnotesize{}S12} & {\footnotesize{}95.5 (0.4)}\tabularnewline
{\footnotesize{}Y10} & {\footnotesize{}78.3 (1.0)} &  & {\footnotesize{}$\mathrm{J11}^{2}$} & \textbf{\textcolor{black}{\footnotesize{}87.4 (0.9)}} &  & {\footnotesize{}L10} & \textbf{\textcolor{black}{\footnotesize{}59.3 (0.9)}}\tabularnewline
{\footnotesize{}C11} & {\footnotesize{}77.1 (0.9)} &  & {\footnotesize{}Y10} & {\footnotesize{}90.2 (0.6)} &  &  & \tabularnewline
{\footnotesize{}L10} & \textbf{\textcolor{black}{\footnotesize{}72.4 (0.9)}} &  & {\footnotesize{}L10} & {\footnotesize{}92.2 (0.6)} &  &  & \tabularnewline
\bottomrule
\end{tabular}
\par\end{centering}{\footnotesize \par}

\end{table}

\subsubsection{Third perspective: traditional model fitting}

For the third perspective, we investigate the AIC, BIC, and deviance
measures. When comparing models fitted by maximum likelihood to the
same data, the smaller the AIC or BIC value is, the better the model
fits, from their respective stand points.

\begin{table}
\caption{Estimated AIC, BIC and deviance for \texttt{Colon}, \texttt{Leukemia} and \texttt{Prostate}.  \label{tab: realdata-Colon-AICBIC}}
\begin{centering}
{\footnotesize{}}%
\begin{tabular}{p{1.3cm} p{0.5cm}p{0.5cm}p{0.5cm}p{0.1cm}p{0.5cm}p{0.5cm}p{0.5cm} p{0.5cm}p{0.1cm}p{0.5cm}p{0.7cm}p{0.7cm}p{0.7cm}}
\toprule 
\multicolumn{4}{c}{\texttt{\footnotesize{}Colon}} &  & \multicolumn{4}{c}{\texttt{\footnotesize{}Leukemia}} &  & \multicolumn{4}{c}{\texttt{\footnotesize{}Prostate}}\tabularnewline
\cmidrule{1-4} \cmidrule{6-9} \cmidrule{11-14} 
 & {\footnotesize{}AIC} & {\footnotesize{}BIC} & {\footnotesize{}Dev.} &  &  & {\footnotesize{}AIC} & {\footnotesize{}BIC} & {\footnotesize{}Dev.} &  &  & {\footnotesize{}AIC} & {\footnotesize{}BIC} & {\footnotesize{}Dev.}\tabularnewline
{\footnotesize{}Lasso} & {\footnotesize{}26.0} & {\footnotesize{}53.6} & {\footnotesize{}0.0} &  &  & {\footnotesize{}56.0} & {\footnotesize{}119.7} & {\footnotesize{}0.0} &  &  & \textcolor{black}{\footnotesize{}62.0} & \textcolor{black}{\footnotesize{}143.3} & {\footnotesize{}0}\textcolor{black}{\footnotesize{}.0}\tabularnewline
\multirow{1}{*}{{\footnotesize{}AdLasso}} & {\footnotesize{}34.9} & {\footnotesize{}49.8} & {\footnotesize{}20.9} &  &  & {\footnotesize{}12.0} & {\footnotesize{}25.6} & {\footnotesize{}0.0} &  &  & \textcolor{black}{\footnotesize{}22.0} & \textcolor{black}{\footnotesize{}50.8} & {\footnotesize{}0}\textcolor{black}{\footnotesize{}.0}\tabularnewline
\multirow{1}{*}{{\footnotesize{}MCP}} & {\footnotesize{}32.1} & {\footnotesize{}44.9} & {\footnotesize{}20.1} &  &  & {\footnotesize{}16.0} & {\footnotesize{}34.2} & {\footnotesize{}0.0} &  &  & \textcolor{black}{\footnotesize{}16.0} & \textcolor{black}{\footnotesize{}36.9} & {\footnotesize{}0}\textcolor{black}{\footnotesize{}.0}\tabularnewline
\multirow{1}{*}{{\footnotesize{}SCAD}} & {\footnotesize{}26.0} & {\footnotesize{}53.6} & {\footnotesize{}0.0} &  &  & {\footnotesize{}48.0} & {\footnotesize{}102.6} & {\footnotesize{}0.0} &  &  & \textcolor{black}{\footnotesize{}38.0} & \textcolor{black}{\footnotesize{}87.8} & {\footnotesize{}0}\textcolor{black}{\footnotesize{}.0}\tabularnewline
\multirow{1}{*}{{\footnotesize{}ImpS}} & {\footnotesize{}35.5} & {\footnotesize{}44.1} & {\footnotesize{}27.5} &  &  & {\footnotesize{}8.0} & {\footnotesize{}17.1} & {\footnotesize{}0.0} &  &  & \textcolor{black}{\footnotesize{}12.0} & \textcolor{black}{\footnotesize{}27.7} & {\footnotesize{}9.4}\tabularnewline
\multirow{1}{*}{{\footnotesize{}L11}} & {\footnotesize{}51.4} & {\footnotesize{}70.5} & {\footnotesize{}33.4} &  & {\footnotesize{}J$11^{1}$} & \textbf{\textcolor{black}{\footnotesize{}20.0}} & \textbf{\textcolor{black}{\footnotesize{}42.7}} & \textbf{\textcolor{black}{\footnotesize{}0.0}} &  & \textcolor{black}{\footnotesize{}S12} & \textcolor{black}{\footnotesize{}36.1} & \textcolor{black}{\footnotesize{}49.2} & {\footnotesize{}26.1}\tabularnewline
\multirow{1}{*}{{\footnotesize{}Y10}} & {\footnotesize{}40.0} & {\footnotesize{}82.5} & {\footnotesize{}0.0} &  & {\footnotesize{}J$11^{2}$} & \textbf{\textcolor{black}{\footnotesize{}18.0}} & \textbf{\textcolor{black}{\footnotesize{}38.4}} & \textbf{\textcolor{black}{\footnotesize{}0.0}} &  & \textcolor{black}{\footnotesize{}L10} & \textbf{\textcolor{black}{\footnotesize{}140.1}} & \textbf{\textcolor{black}{\footnotesize{}158.5}} & \textbf{\textcolor{black}{\footnotesize{}126.1}}\tabularnewline
\multirow{1}{*}{{\footnotesize{}C11}} & {\footnotesize{}45.2} & {\footnotesize{}68.6} & {\footnotesize{}23.2} &  & {\footnotesize{}Y10} & {\footnotesize{}38.0} & {\footnotesize{}81.2} & {\footnotesize{}0.0} &  &  &  &  & \tabularnewline
\multirow{1}{*}{{\footnotesize{}L10}} & \textbf{\textcolor{black}{\footnotesize{}48.6}} & \textbf{\textcolor{black}{\footnotesize{}63.5}} & \textbf{\textcolor{black}{\footnotesize{}34.6}} &  & {\footnotesize{}L10} & {\footnotesize{}10.0} & {\footnotesize{}21.3} & {\footnotesize{}0.0} &  &  &  &  & \tabularnewline
\bottomrule
\end{tabular}

\par\end{centering}{\footnotesize \par}

\end{table}

From Table \ref{tab: realdata-Colon-AICBIC}, the model for \texttt{Colon} with zero
$\widehat{F}$ and $\widehat{G}$ values also has relatively large
AIC, BIC and deviance values (bolded in the Table) compared to the models with large $\widehat{F}$
and $\widehat{G}$ values. The results are similar for the other two data sets, except that the deviance values for \texttt{Leukemia} are extremely small due to easy classification
nature of the data.\\

In summary, we see that the low (near zero) $\widehat{F}$ and $\widehat{G}$
values for the investigated sets of selected genes above are supported
from the three perspectives. Our PAVI approach provides a valid tool
for checking the reliability and reproducibility of a given set of
selected variables when the true model is not known. To be fair, we want to emphasize that the poor $\widehat{F}$ and $\widehat{G}$ values of some of the selection methods are based on the logistic regression perspective, although Table \ref{tab: Logistic}
seems to suggest that logistic regression works at least as well as SVM.

\section{CONCLUSION}

There are many variable selection methods, but so far most of investigations
on their behaviors are limited to theoretical studies and somewhat
scattered simulation results, which may have little to do with a specific
dataset at hand. There is a severe lack of valid performance measures
that are computable based on data alone. This leads to the pessimistic
view that ``For real data, nothing can be said strongly about which method
is better for describing the data generation mechanism since no one knows the truth.'' Sound implementable variable
selection diagnostic tools can shed a positive light on the
matter. 

\citet{nanying} proposed an approach to gain insight on how many
variables are likely missed and how many are not quite justifiable
for an outcome of a variable selection process. In real applications,
it is often of interest and important to summarize the two types of
selection errors into a single measure to characterize the behavior
of a variable selection method. Due to this reason, $F$- and $G$-measures
are gaining popularity in model selection literature. If we are given
a data set for which several model selection methods are considered,
prior to this work, the available model diagnostic tools can only
tell us (a) Which methods are more unstable; (b) How many terms are
likely missed or unsupported. This information, unlike the $F$- and
$G$-measures, may not be enough to give one a good sense of the overall
model selection performance. In this paper, we have advanced the line
of research on model selection diagnostics by providing a valid estimation
of $F$- and $G$-measures.

We have proved that the estimated $F$- and $G$-measures are uniformly
consistent as long as the weighting is weakly consistent. The simulation
results clearly show that the $\widehat{F}$ and $\widehat{G}$ based
on our PAVI approach nicely characterizes the overall performance
of the model selection outcomes. The information can be utilized for
comparing different methods for the data at hand. 

We have used three real data examples to demonstrate the utility of
our PAVI methodology. There have been many variable selection results
reported in the literature on these data sets. A careful study with
multiple perspectives has provided strong evidence to suggest that
some of the variable selection outcomes may be far away from the best
set of variables to use for logistic regression or SVM with the given
information.

\bibliographystyle{rss}
\bibliography{ref}

\section*{Appendix for ``Performance Assessment of High-dimensional Variable
Identification''}

In this appendix we provide technical proofs for the theorems and
lemmas in ``Performance Assessment of High-dimensional Variable Identification''.

\section*{Proof of Theorem \ref{thm:f-measure}}

\subsection*{Part I: $F$-measure}
\begin{proof}
Denote by $\nabla$ the symmetric difference between two sets. Estimated
$F$-measure can be rewritten as
\[
\widehat{F}(\az)=\sum w_{k}F(\az;\ak),\qquad F(\az;\ak)=\frac{|\az|+|\ak|-|\az\nabla\ak|}{|\az|+|\ak|}.
\]
We have
\begin{align*}
|\widehat{F}(\mathcal{A}^{0})-F(\az)| & =\left|\sum w_{k}F(\az;\ak)-F(\az)\right|\\
 & =\left|\sum w_{k}(F(\az;\ak)-F(\az))\right|\le\sum w_{k}|F(\az;\ak)-F(\az)|\\
 & =\sum w_{k}\left|1-\frac{|\az\nabla\ak|}{|\az|+|\ak|}-1+\frac{|\az\nabla\as|}{|\az|+|\as|}\right|\\
 & =\sum w_{k}\left|\frac{|\az|\cdot(|\az\nabla\as|-|\az\nabla\ak|)+|\ak|\cdot|\az\nabla\as|-|\as|\cdot|\az\nabla\ak|}{(|\az|+|\ak|)(|\az|+|\as|)}\right|\\
 & \leq\text{\ensuremath{\underbrace{\sum w_{k}\frac{|\az|\cdot||\az\nabla\as|-|\az\nabla\ak||}{(|\az|+|\ak|)(|\az|+|\as|)}}_{A}}+\ensuremath{\underbrace{\sum w_{k}\frac{|\ak|\cdot||\az\nabla\as|-|\az\nabla\ak||}{(|\az|+|\ak|)(|\az|+|\as|)}}_{B}}}\\
 & +\underbrace{\sum w_{k}\frac{||\ak|-|\as||\cdot|\az\nabla\ak|}{(|\az|+|\ak|)(|\az|+|\as|)}}_{C}.
\end{align*}
For ease of notation, we divide the right-most hand side of the above
inequality into three parts and denote them by $A$, $B$, and $C$
respectively. Note that since $\left||\az\nabla\as|-|\az\nabla\ak|\right|\leq|\as\nabla\ak|$,
we have 
\[
A\leq\sum w_{k}\frac{|\az|\cdot|\as\nabla\ak|}{(|\az|+|\ak|)(|\az|+|\as|)}\leq\sum w_{k}\frac{|\as\nabla\ak|}{|\as|}.
\]
Similarly, it can be shown that 
\[
B\leq\sum w_{k}\frac{|\as\nabla\ak|}{|\as|}.
\]
Let us now prove a similar bound also holds for $C.$ Specifically,
we have
\begin{align*}
C & =\sum w_{k}\frac{||\ak|-|\as||\cdot|\az\nabla\ak|}{(|\az|+|\ak|)(|\az|+|\as|)}\leq\sum w_{k}\frac{\left||\ak|-|\as|\right|}{|\az|+|\as|}\\
 & =\sum w_{k}\frac{\left|(|\ak\backslash\as|+|\ak\cap\as|)-(|\as\backslash\ak|+|\ak\cap\as|)\right|}{|\az|+|\as|}\\
 & =\sum w_{k}\frac{\left||\ak\backslash\as|-|\as\backslash\ak|\right|}{|\az|+|\as|}\leq\sum w_{k}\frac{|\ak\backslash\as|+|\as\backslash\ak|}{|\az|+|\as|}\\
 & =\sum w_{k}\frac{|\ak\nabla\as|}{|\az|+|\as|}\leq\sum w_{k}\frac{|\ak\nabla\as|}{|\as|}.
\end{align*}
It follows that for any $\az$ in $\mathbb{C}$
\[
|\widehat{F}(\mathcal{A}^{0})-F(\az)|\leq A+B+C\leq3\sum w_{k}\frac{|\as\nabla\ak|}{|\as|}.
\]
Therefore,
\[
\sup_{\az\in\mathbb{C}}|\widehat{F}(\mathcal{A}^{0})-F(\az)|\leq3\sum w_{k}\frac{|\as\nabla\ak|}{|\as|}.
\]
Now under the assumption that the model weighting $w$ is weakly consistent,
\[
\sum w_{k}\frac{|\as\nabla\ak|}{|\as|}\overset{p}{\rightarrow}0.
\]
We have proved $\sup_{\az\in\mbox{\ensuremath{\mathbb{C}}}}|\widehat{F}(\mathcal{A}^{0})-F(\az)|\overset{p}{\rightarrow}0.$
\end{proof}

\subsection*{Part II: $G$-measure}
\begin{proof}
For a given $\az$ in $\mathbb{C}$, the estimated $G$-measure can
be rewritten as 
\[
\widehat{G}(\mathcal{A}^{0})=\sum w_{k}G(\az;\ak),\qquad G(\az;\ak)=\frac{|\az|+|\ak|-|\az\nabla\ak|}{2\sqrt{|\az|\cdot|\ak|}}.
\]
Suppose $|\widehat{G}(\mathcal{A}^{0})-G(\az)|$ does not converge
to $0$ in probability uniformly over $\mathbb{C}$, then there exist
some subsequence $n_{1},n_{2},\cdots$, $\epsilon_{1}>0,\delta>0$,
$\az_{n_{j}}\in\mathbb{C}$, and sets $\mathcal{S}_{n_{j}}$, s.t.
$P(\mathcal{S}_{n_{j}})\geq\delta$ and $|\widehat{G}(\az_{n_{j}})-G(\az_{n_{j}})|>\epsilon_{1}$
on $\mathcal{S}_{n_{j}}.$ For ease of notation, we denote $\az_{n_{j}}$
as $\az$ in the following proof.

With the above, we first prove that we must have $\text{\ensuremath{\frac{|\az|}{|\as|}\overset{p}{\rightarrow}}0}$
on $\mathcal{S}_{n_{j}}$ as $n_{j}\rightarrow\infty$. If not, then
there exist $\epsilon_{2}>0,$ a subsequence $n_{j_{l}}$ and sets
$\mathcal{N}_{n_{j_{l}}}$ such that on $\mathcal{N}_{n_{j_{l}}}$
we have $\frac{|\az|}{|\as|}>\epsilon_{2}>0$. Then we can actually
prove $|\widehat{G}(\mathcal{A}^{0})-G(\az)|\overset{p}{\longrightarrow}0$
on $\mathcal{N}_{n_{j_{l}}}$ as follows.

By definition of $\widehat{G}$ and $G$, and $\frac{|\az|}{|\as|}>\epsilon_{2}>0$
on $\mathcal{N}_{n_{j_{l}}}$, we have
\begin{align*}
|\widehat{G}(\mathcal{A}^{0})-G(\az)| & =|\sum w_{k}G(\az;\ak)-G(\az)|\\
 & \leq\sum w_{k}|G(\az;\ak)-G(\az)|\\
 & =\sum w_{k}\left|\frac{|\az|+|\ak|-|\az\nabla\ak|}{2\sqrt{|\az|\cdot|\ak|}}-\frac{|\az|+|\as|-|\az\nabla\as|}{2\sqrt{|\az|\cdot|\as|}}\right|\\
 & \leq\sum w_{k}\frac{|\sqrt{|\as|}-\sqrt{|\ak|}|\cdot||\az|+|\ak|-|\az\nabla\ak||}{2\sqrt{|\as|\cdot|\az|\cdot|\ak|}}\\
 & +\sum w_{k}\frac{\sqrt{|\ak|}\cdot||\ak|-|\as|+|\az\nabla\as|-|\az\nabla\ak||}{2\sqrt{|\as|\cdot|\az|\cdot|\ak|}}\\
 & \leq\underbrace{\sum w_{k}\frac{|\sqrt{|\as|}-\sqrt{|\ak|}|\cdot||\az|+|\ak|-|\az\nabla\ak||}{2\sqrt{|\as|\cdot|\az|\cdot|\ak|}}}_{A}\\
 & +\underbrace{\sum w_{k}\frac{||\ak|-|\as||}{2\sqrt{|\as|\cdot|\az|}}}_{B}+\underbrace{\sum w_{k}\frac{||\az\nabla\as|-|\az\nabla\ak||}{2\sqrt{|\as|\cdot|\az|}}}_{C}.
\end{align*}
For notational convenience, we divide the right-most-hand side of
the above inequality into three parts and denote them by $A$, $B$,
and $C$ respectively. For part $A$, because $|\az|+|\ak|-|\az\nabla\ak|=2|\az\cap\ak|$
and $\left||\as|-|\ak|\right|\leq|\as\nabla\ak|$, together with $|\az\cap\ak|\leq\sqrt{|\az|\cdot|\ak|}$,
we have
\[
A=\sum w_{k}\frac{\left||\as|-|\ak|\right|\cdot|\az\cap\ak|}{\left(\sqrt{|\as|}+\sqrt{|\ak|}\right)\sqrt{|\as|\cdot|\az|\cdot|\ak|}}\leq\sum w_{k}\frac{|\as\nabla\ak|}{|\as|}.
\]
For part $B$, since $||\ak|-|\as||\leq|\ak\nabla\as|$ and $\frac{|\az|}{|\as|}>\epsilon_{2}>0$
on $\mathcal{N}_{n_{j_{l}}}$, we have 
\[
B=\sum w_{k}\frac{\left||\ak|-|\as|\right|}{2\sqrt{|\as|\cdot|\az|}}\leq\frac{1}{2\sqrt{\epsilon_{2}}}\sum w_{k}\frac{|\ak\nabla\as|}{|\as|}.
\]
For part $C$, it follows from the facts that $||\az\nabla\as|-|\az\nabla\ak||\leq|\as\nabla\ak|$
and that $\frac{|\az|}{|\as|}>\epsilon_{2}>0$ on $\mathcal{N}_{n_{j_{l}}}$,
we have 
\[
C=\sum w_{k}\frac{||\az\nabla\as|-|\az\nabla\ak||}{2\sqrt{|\as|\cdot|\az|}}\leq\frac{1}{2\sqrt{\epsilon_{2}}}\frac{\sum w_{k}|\as\nabla\ak|}{|\as|}.
\]
Consequently, we have that on $\mathcal{N}_{n_{j_{l}}}$, 
\[
|\widehat{G}(\mathcal{A}^{0})-G(\az)|\leq A+B+C\leq(1+\frac{1}{\sqrt{\epsilon_{2}}})\sum w_{k}\frac{|\as\nabla\ak|}{|\as|}.
\]
Under the assumption that the model weighting $w$ is weakly consistent,
\[
\sum w_{k}\frac{|\as\nabla\ak|}{|\as|}\overset{p}{\rightarrow}0,
\]
 we must have $|\widehat{G}(\mathcal{A}^{0})-G(\az)|\overset{p}{\rightarrow}0$
on $\mathcal{N}_{n_{j_{l}}}$. This contradicts with the statement
that $|\widehat{G}(\mathcal{A}^{0})-G(\az)|>\epsilon_{1}>0$ on $\mathcal{S}_{n_{j}}$.
Therefore, we have proved that $\frac{|\az|}{|\as|}\overset{p}{\longrightarrow}0$
on $\mathcal{S}_{n_{j}}$ under the beginning supposition.

Next, we prove actually we must have $|\widehat{G}(\mathcal{A}^{0})-G(\az)|\overset{p}{\rightarrow}0$
on $\mathcal{S}_{n_{j}}$ as $n_{j}\rightarrow\infty$. Because $\frac{|\az|}{|\as|}\overset{p}{\rightarrow}0$
on $\mathcal{S}_{n_{j}}$, we can set $\delta_{n}=\sqrt{\frac{|\az|}{|\as|}}$,
then $\delta_{n}\overset{p}{\rightarrow}0$ and $\frac{|\az|}{|\as|\cdot\delta_{n}}=\delta_{n}\overset{p}{\rightarrow}0.$
Then
\[
|G(\az)|=\frac{||\az|+|\as|-|\az\nabla\as||}{2\sqrt{|\as|\cdot|\az|}}=\frac{|\az\cap\as|}{\sqrt{|\az|\cdot|\as|}}\leq\sqrt{\frac{|\az|}{|\as|}}\overset{p}{\rightarrow}0.
\]
That is $G(\az)\overset{p}{\rightarrow}0$. Now we prove that we also
have $\widehat{G}(\mathcal{A}^{0})\overset{p}{\rightarrow}0$ as follows.
Observe on $\mathcal{S}_{n_{j}}$
\begin{align*}
\widehat{G}(\mathcal{A}^{0}) & =\sum I(|\ak|\leq|\as|\cdot\delta_{n})\cdot w_{k}\frac{|\az\cap\ak|}{\sqrt{|\az|\cdot|\ak|}}+\sum I(|\ak|>|\as|\cdot\delta_{n})\cdot w_{k}\frac{|\az\cap\ak|}{\sqrt{|\az|\cdot|\ak|}}\\
 & \leq\sum I(|\ak|\leq|\as|\cdot\delta_{n})\cdot w_{k}+\sum I(|\ak|>|\as|\cdot\delta_{n})\cdot w_{k}\frac{|\az\cap\ak|}{\sqrt{|\az|\cdot|\ak|}}.
\end{align*}
Then because $\sum w_{k}\frac{|\ak\nabla\as|}{|\as|}\overset{p}{\rightarrow}0$
and
\begin{align*}
\sum w_{k}\frac{|\ak\nabla\as|}{|\as|} & \geq\sum w_{k}\frac{||\as|-|\ak||}{|\as|}\\
 & \geq\sum w_{k}\frac{||\as|-|\ak||}{|\as|}\cdot I(|\ak|\leq|\as|\cdot\delta_{n})\\
 & \geq\frac{1}{2}\sum w_{k}\cdot I(|\ak|\leq|\as|\cdot\delta_{n}).
\end{align*}
We know $\sum I(|\ak|\leq|\as|\cdot\delta_{n})\cdot w_{k}\overset{p}{\rightarrow}0$.
On $\mathcal{S}_{n_{j}},$ we also have
\begin{align*}
 & \sum I(|\ak|>|\as|\cdot\delta_{n})\cdot w_{k}\frac{|\az\cap\ak|}{\sqrt{|\az|\cdot|\ak|}}\\
\leq & \sum I(|\ak|>|\as|\cdot\delta_{n})\cdot w_{k}\sqrt{\frac{|\az|}{|\ak|}}\\
\leq & \sum I(|\ak|>|\as|\cdot\delta_{n})\cdot w_{k}\sqrt{\frac{|\az|}{|\as|\cdot\delta_{n}}}\\
\overset{p}{\rightarrow} & 0,
\end{align*}
since $\frac{|\az|}{|\as|\cdot\delta_{n}}\overset{p}{\rightarrow}0$
on $\mathcal{S}_{n_{j}}$. Therefore, we have shown $\widehat{G}(\mathcal{A}^{0})\overset{p}{\rightarrow}0$
on $\mathcal{S}_{n_{j}}$.

Now since we have proved that on $\mathcal{S}_{n_{j}}$, $G(\az)\overset{p}{\rightarrow}0$
and $\widehat{G}(\mathcal{A}^{0})\overset{p}{\rightarrow}0$, so $|\widehat{G}(\mathcal{A}^{0})-G(\az)|\overset{p}{\rightarrow}0$
on $\mathcal{S}_{n_{j}}$, which contradicts with the beginning supposition
that $|\widehat{G}(\mathcal{A}^{0})-G(\az)|>\epsilon_{1}>0$ on $\mathcal{S}_{n_{j}}$.
Therefore the supposition does not hold, and we have proved the $|\widehat{G}(\mathcal{A}^{0})-G(\az)|$
does converge to $0$ in probability uniformly over $\mathbb{C}$.
\end{proof}

\section*{Proof of Theorem \ref{thm:sd-f-measure}}

\subsection*{Part I: standard deviation of $F$-measure}
\begin{proof}
For any $\az$ in $\mathbb{C}$, by definition of the standard deviation
of $F$-measure, we have 
\begin{align*}
\mathrm{sd}\big(\widehat{F}(\mathcal{A}^{0})\big) & \equiv\sqrt{\sum w_{k}\big(F(\az;\ak)-\widehat{F}(\mathcal{A}^{0})\big)^{2}}\\
\leq & \sqrt{\sum w_{k}|F(\az;\ak)-\widehat{F}(\mathcal{A}^{0})|}\\
\leq & \sqrt{\sum w_{k}|F(\az;\ak)-F(\mathcal{A}^{0})|+|F(\az)-\widehat{F}(\mathcal{A}^{0})|}.
\end{align*}
Using the facts proved in the proof for Theorem \ref{thm:f-measure},
\[
|\widehat{F}(\mathcal{A}^{0})-F(\az)|\le\sum w_{k}|F(\az;\ak)-F(\az)|\leq3\sum w_{k}\frac{|\as\nabla\ak|}{|\as|},
\]
we know
\begin{align*}
\text{sd}(\widehat{F}(\mathcal{A}^{0})) & \leq\sqrt{6\sum w_{k}\frac{|\as\nabla\ak|}{|\as|}},
\end{align*}
and
\[
\sup_{\az\in\mathbb{C}}\text{sd}(\widehat{F}(\mathcal{A}^{0}))\leq\sqrt{6\sum w_{k}\frac{|\as\nabla\ak|}{|\as|}}\overset{p}{\rightarrow}0
\]
under the assumption that the model weighting $w$ is weakly consistent.
\end{proof}

\subsection*{Part II: standard deviation of $G$-measure}
\begin{proof}
For any $\az$ in $\mathbb{C}$, by definition of the standard deviation
of $G$-measure, we have
\begin{align*}
\mathrm{sd}\big(\widehat{G}(\mathcal{A}^{0})\big) & \equiv\sqrt{\sum w_{k}\big(G(\az;\ak)-\widehat{G}(\mathcal{A}^{0})\big)^{2}}\\
\leq & \sqrt{\sum w_{k}|G(\az;\ak)-\widehat{G}(\mathcal{A}^{0})|}\\
\leq & \sqrt{\sum w_{k}|G(\az;\ak)-G(\mathcal{A}^{0})|+|G(\az)-\widehat{G}(\mathcal{A}^{0})|}.
\end{align*}
Using the facts in Theorem \ref{thm:f-measure}, we have
\[
|\widehat{G}(\mathcal{A}^{0})-G(\az)|\overset{p}{\rightarrow}0.
\]
So it suffices to show $\sum w_{k}|G(\az;\ak)-G(\mathcal{A}^{0})|\overset{p}{\rightarrow}0$.
The arguments are similar to those in the proof of Theorem \ref{thm:f-measure}.
For completeness, the full proof is given below. 

Suppose $\sum w_{k}|G(\az;\ak)-G(\mathcal{A}^{0})|$ does not converge
to $0$ in probability uniformly over $\mathbb{C}$, then there exist
some subsequence $n_{1},n_{2},\cdots$, $\epsilon_{1}>0,\delta>0$,
$\az_{n_{j}}\in\mathbb{C}$, and sets $\mathcal{S}_{n_{j}}$, s.t.
$P(\mathcal{S}_{n_{j}})\geq\delta$ and $\sum w_{k}|G(\az_{n_{j}};\ak)-G(\mathcal{A}_{n_{j}}^{0})|>\epsilon_{1}$
on $\mathcal{S}_{n_{j}}.$ For ease of notation, we denote $\az_{n_{j}}$
as $\az$. We first prove that we must have $\text{\ensuremath{\frac{|\az|}{|\as|}\overset{p}{\rightarrow}}0}$
on $\mathcal{S}_{n_{j}}$ as $n_{j}\rightarrow\infty$. If not, then
there exist $\epsilon_{2}>0,$ a subsequence $n_{j_{l}}$ and sets
$\mathcal{N}_{n_{j_{l}}}$ such that on $\mathcal{N}_{n_{j_{l}}}$
we have $\frac{|\az|}{|\as|}>\epsilon_{2}>0$. Then we can actually
prove $\sum w_{k}|G(\az;\ak)-G(\mathcal{A}^{0})|\overset{p}{\longrightarrow}0$
on $\mathcal{N}_{n_{j_{l}}}$ as follows. On $\mathcal{N}_{n_{j_{l}}}$,
since $\frac{|\az|}{|\as|}>\epsilon_{2}>0$ , we have that

\begin{align*}
\sum w_{k}|G(\az;\ak)-G(\az)| & \leq\underbrace{\sum w_{k}\frac{|\sqrt{|\as|}-\sqrt{|\ak|}|\cdot||\az|+|\ak|-|\az\nabla\ak||}{2\sqrt{|\as|\cdot|\az|\cdot|\ak|}}}_{A}\\
 & +\underbrace{\sum w_{k}\frac{||\ak|-|\as||}{2\sqrt{|\as|\cdot|\az|}}}_{B}+\underbrace{\sum w_{k}\frac{||\az\nabla\as|-|\az\nabla\ak||}{2\sqrt{|\as|\cdot|\az|}}}_{C}\\
\leq & (1+\frac{1}{\sqrt{\epsilon_{2}}})\sum w_{k}\frac{|\as\nabla\ak|}{|\as|}.
\end{align*}
Under the assumption that the model weighting $w$ is weakly consistent,
\[
\sum w_{k}\frac{|\as\nabla\ak|}{|\as|}\overset{p}{\rightarrow}0,
\]
we must have $\sum w_{k}|G(\az;\ak)-G(\mathcal{A}^{0})|\overset{p}{\rightarrow}0$
on $\mathcal{N}_{n_{j_{l}}}$. This contradicts with the statement
that $\sum w_{k}|G(\az;\ak)-G(\mathcal{A}^{0})|>\epsilon_{1}>0$ on
$\mathcal{S}_{n_{j}}$. Therefore, we have proved that $\frac{|\az|}{|\as|}\overset{p}{\longrightarrow}0$
on $\mathcal{S}_{n_{j}}$ under the beginning supposition.

Next, we prove actually we must have $\sum w_{k}|G(\az;\ak)-G(\mathcal{A}^{0})|\overset{p}{\rightarrow}0$
on $\mathcal{S}_{n_{j}}$ as $n_{j}\rightarrow\infty$. Similar to
the proof in Theorem \ref{thm:f-measure}, we can prove that $G(\az)\overset{p}{\rightarrow}0$
and $\widehat{G}(\mathcal{A}^{0})\overset{p}{\rightarrow}0$ on $\mathcal{S}_{n_{j}}$.
We then have 
\[
\sum w_{k}|G(\az;\ak)-G(\mathcal{A}^{0})|\leq\sum w_{k}G(\az;\ak)+G(\az)=\widehat{G}(\mathcal{A}^{0})+G(\az)\overset{p}{\rightarrow}0
\]
on $\mathcal{S}_{n_{j}}$, which contradicts with the beginning supposition
that $\sum w_{k}|G(\az_{n_{j}};\ak)-G(\mathcal{A}_{n_{j}}^{0})|>\epsilon_{1}>0$
on $\mathcal{S}_{n_{j}}$. Therefore the supposition does not hold,
and we have proved the $\sum w_{k}|G(\az_{n_{j}};\ak)-G(\mathcal{A}_{n_{j}}^{0})|$
does converge to $0$ in probability uniformly over $\mathbb{C}$.
Since we have $\mathrm{sd}\big(\widehat{G}(\mathcal{A}^{0})\big)\leq\sqrt{\sum w_{k}|G(\az;\ak)-G(\mathcal{A}^{0})|+|G(\az)-\widehat{G}(\mathcal{A}^{0})|}\overset{p}{\rightarrow}0$
for any $\az\in\mathbb{C}$, we have proved
\[
\sup_{\az\in\mathbb{C}}|\mathrm{sd}\big(\widehat{G}(\mathcal{A}^{0})\big)|\overset{p}{\longrightarrow}0\qquad\mbox{as}\ n\rightarrow\infty.
\]

\end{proof}

\end{document}